\documentclass[a4paper,12pt]{article}

\usepackage{amsmath,amssymb,amsfonts}
\usepackage{epsfig,cancel,cite}

\newlength{\xtrawidth}
\newlength{\xtraheight}
\newcommand{\beq}{\begin{equation}}
\newcommand{\eeq}{\end{equation}}
\newcommand{\ba}{\begin{array}}
\newcommand{\ea}{\end{array}}
\newcommand{\bea}{\begin{eqnarray}}
\newcommand{\eea}{\end{eqnarray}}
\newcommand{\bean}{\begin{eqnarray*}}
\newcommand{\eean}{\end{eqnarray*}}
\newcommand{\eref}[1]{(\ref{#1})}

\newcommand{\nn}{\nonumber}
\newcommand{\comment}[1]{}

\newcommand{\rk}{\mathop{{\rm rk}}}
\newcommand{\coker}{\mathop{{\rm coker}}}
\newcommand{\ch}{{\rm ch}}
\newcommand{\td}{{\rm Td}}
\newcommand{\ind}{\mathop{{\rm ind}}}

\newcommand{\IP}{\mathbb{P}}

\newcommand{\II}{\mathbb{I}}
\newcommand{\IZ}{\mathbb{Z}}
\newcommand{\cO}{{\cal O}}
\newcommand{\cN}{{\cal N}}
\newcommand{\cF}{{\cal F}}
\newcommand{\cA}{{\cal A}}
\newcommand{\cB}{{\cal B}}
\newcommand{\cC}{{\cal C}}
\newcommand{\cL}{{\cal L}}
\newcommand{\cV}{{\cal V}}
\newcommand{\sumi}{\sum\limits_{i=1}^{r_B}}
\newcommand{\sumj}{\sum\limits_{j=1}^{r_C}}

\def\fnote#1#2{\begingroup\def\thefootnote{#1}\footnote{#2}
     \addtocounter{footnote}{-1}\endgroup}

\newcommand{\sseq}[3]{0 \to #1 \to #2 \to #3 \to 0}

\newcommand{\tconf}[1]{{\tiny \left[\begin{matrix}#1\end{matrix}\right]}}

\newcommand{\setall}{\setcounter{equation}{0}
        \setcounter{theorem}{0}
}

\begin{document}

\title{{\LARGE
Monad Bundles in Heterotic String Compactifications
}}
\author{
Lara Anderson${}^{1,3}$,
Yang-Hui He${}^{1,2,3}$,
Andr\'e Lukas${}^{3}$
}
\date{}
\maketitle
\begin{center}
{\small
${}^1$ {\it Mathematical Institute, Oxford University, 
\\ $~~~~~$ 24-29 St.~Giles', Oxford OX1 3LB, U.K.}\\[0.2cm]
${}^2${\it Merton College, Oxford, OX1 4JD, U.K.}\\[0.2cm]
${}^3${\it Rudolf Peierls Centre for Theoretical Physics, Oxford
  University,\\
$~~~~~$ 1 Keble Road, Oxford, OX1 3NP, U.K.}\\
\fnote{}{anderson@maths.ox.ac.uk}
\fnote{}{hey@maths.ox.ac.uk} 
\fnote{}{lukas@physics.ox.ac.uk}
}
\end{center}

\abstract{In this paper, we study positive monad vector bundles on complete intersection Calabi-Yau manifolds in the context of $E_8\times E_8$ heterotic string compactifications. We show that the class of such bundles, subject to the heterotic anomaly condition, is finite and consists of about $7000$ models. We explain how to compute the complete particle spectrum for these models. In particular, we prove the absence of vector-like family anti-family pairs in all cases. We also verify a set of highly non-trivial necessary conditions for the stability of the bundles. A full stability proof will appear in a companion paper. A scan over all models shows that even a few rudimentary physical constraints reduces the number of viable models drastically.}

\newpage

\tableofcontents

%
%
\section{Introduction}
\setall
As string and M-theory continue to develop, it remains a problem of central 
importance to produce models that are relevant to 4-dimensional particle phenomenology. While many approaches to this goal have been explored over the years, a string model with exactly the particle content and detailed properties of the standard model remains elusive. One of the first and currently most successful approaches to this challenge has been provided by heterotic string theory. Because they naturally incorporate gauge unification, heterotic models are particularly well suited for use in string phenomenology. The vector bundles with ${\rm SU}(n)$ structure group used in heterotic models lead to the gauge groups of grand unified theories (GUTs) in 4-dimensions and under suitable symmetry breaking (that is, Wilson lines, etc) can 
contain the symmetry of the standard model. More specifically, compactification of
10-dimensional heterotic string theory on Calabi-Yau three-folds equipped with (poly-)stable holomorphic ${\rm SU}(n)$ vector bundles  leads to $N=1$ supersymmetric versions of GUTs.

Despite substantial recent progress \cite{Donagi:2004ia,Braun:2005ux,Braun:2005nv,Bouchard:2005ag,Donagi:2004su,Braun:2005zv,Distler:2007av,Blumenhagen:2006wj,Bouchard:2008bg,Candelas:2007ac,Donagi:2004qk},
heterotic model-building continues to present a number of formidable mathematical obstacles. In addition to a Calabi-Yau three-fold $X$, heterotic models require two holomorphic (poly-)stable vector bundles $V$ and $\tilde{V}$. Except for the simplest case of the so-called ``standard embedding'' (in which $V$ is taken to be the tangent bundle to the Calabi-Yau and $\tilde{V}$ is trivial) explicit constructions of both the Calabi-Yau three-fold and the vector bundle, $V$ are generally hard to obtain and difficult to analyze mathematically. It is our goal in this work to present techniques studying a large class of heterotic models in detail. We utilize the well-known monad construction of vector bundles to build bundles over the set of compete intersection Calabi-Yau manifolds.

It is our hope that by formulating a systematic construction of a large class of vector bundles over an explicit and relatively simple set of Calabi-Yau manifolds, we can build a substantial number of heterotic models which can be thoroughly scanned for physically relevant properties. This program was begun in \cite{Anderson:2007nc} in which we laid out an algorithmic approach to bundle constructions over cyclic Calabi-Yau three-folds defined as complete intersections in a single projective space. In this work, we greatly extend our class of bundles and manifolds by generalizing the techniques to Calabi-Yau manifolds obtained as complete intersections in products of (un-weighted) projective spaces. From the 7890 such complete intersection Calabi-Yau manifolds (CICYs) classified in~\cite{Candelas:1987kf,Candelas:1987du,Green:1987cr,Green:1988,He:1990pg,Gagnon:1994ek}, we consider the 4500 or so ``favourable"  ones, by which we mean CICYs whose second cohomology entirely descends from the ambient space. In this paper, we focus on the ``traditional" class of positive monads, that is, monads defined using strictly positive line bundles only. In a forthcoming publication~\cite{zeropaper} we will show that this condition of positivity can, in fact, be somewhat relaxed.

This paper has two main objectives. First, we will show that there is a finite number of positive monads bundles on favourable CICYs and provide a complete classification. Second, we will develop algorithms to calculate the complete particle spectrum of all such monads and apply these methods to carry out a statistical analysis and identify promising particle physics models. 
\comment{We also derive an essential technical result underlying our discussion, namely a systematic method to calculate the cohomology of all line bundles on favourable CICYs.} Finally, we perform a number of checks for the stability of positive monad bundles. A systematic proof of stability will be presented in the companion paper~\cite{stabpaper}.

The plan of the paper is as follows. In Sections 2, 3 and 4 we briefly review some general facts about heterotic model building, complete intersection Calabi-Yau manifolds and the monad construction, respectively. Section 5 summarises the various physical and mathematical constraints on positive monads and why the number of such monads is finite; we then present a complete classification. 
Some non-trivial checks for the stability of these bundles are carried out in Section 6. Computation of the particle spectrum is discussed in Section 7, before we conclude in Section 8. To simplify the discussion in the main body of the paper, many of the underlying mathematical methods and technical results have been collected in appendices. Appendix A summarises our notation and conventions throughout the paper. In Appendix B we review some mathematical methods \comment{notably Koszul resolutions and Leray spectral sequences and show how the ranks of maps in Leray  sequences can be computed based on the Bott-Borel-Weil theorem. These results facilitate the computation of line bundle cohomology on CICYs.}  and formulae which are essential for our calculations. 
Appendix C collects technical results on CICYs, most of them well-known, some, such as the identification of redundancies in the CICY list, new. 

\section{Heterotic Calabi-Yau Model Building}\label{s:constraint}
To set the scene,  we start by briefly reviewing the basics facts on $E_8\times E_8$ heterotic Calabi-Yau model building. For a more complete discussion see for example \cite{Candelas:1985en,GSW,Witten:1985bz,Donagi:2004ia}.

A heterotic Calabi-Yau model is specified by four pieces of data, a Calabi-Yau manifolds $X$, the observable and hidden holomorphic vector bundles $V$ and $\tilde{V}$ on X, each with a structure group contained in $E_8$ and a holomorphic curve $C\subset X$ with associated homology class $W=[C]\in H_2(X,\mathbb{Z})$. Physically, the curve $C$ is wrapped by five-branes stretching across the four-dimensional uncompactified space-time. While models without five-branes are of course possible we would like to maintain a general viewpoint and include this possibility. On this data three physical constraints have to be imposed.
\begin{itemize}
 \item Anomaly cancellation: Anomaly cancellation in the heterotic string imposes a topological condition which relates the Calabi-Yau manifold $X$, the two vector bundles and the five-brane class $W$. For the case of bundles $V$ and $\tilde{V}$ with $c_1(V)=c_1(\tilde{V})=0$ it can be written as
 \begin{equation}
  c_2(TX)-c_2(V)-c_2(\tilde{V})=W\; . \label{anomaly}
 \end{equation}
 \item Effectiveness: To ensure four-dimensional $N=1$ supersymmetry the five-brane has to wrap a holomorphic curve. Hence, the five-brane class $W$ must be chosen such that it indeed has a holomorphic curve representative $C$, with $W=[C]$. Classes $W\in H_2(X,\mathbb{Z})$ with this property are called {\em effective}.
 \item Stability:  The Donaldson-Uhlenbeck-Yau theorem~\cite{duy} guarantees the existence of a connection satisfying the hermitian Yang-Mills equations (and, hence, preserving $N=1$ supersymmetry) on a holomorphic vector bundle, provided this bundle is {\em (poly)-stable}. Hence, both $V$ and $\tilde{V}$ must be (poly)-stable holomorphic vector bundles on $X$.
\end{itemize}
Since the notion of stability of a vector bundle is perhaps not overly familiar it is worth providing a definition.  In this paper, we will not be concerned with poly-stability but only with the slightly stronger condition of {\em stability}. To define this condition, one needs to introduce the {\em slope} of a bundle (or coherent sheaf) $V$ by
\begin{equation}\label{slope}
\mu (V)\equiv \frac{1}{\rk(V)}\int_{X}c_{1}(V)\wedge J\wedge J\; ,  
\end{equation}  
where $J$ is the K\"ahler form on $X$. Then, a bundle $V$ is called stable if $\mu
(\cF)<\mu (V)$ for any coherent sub-sheaf $\cF\subset V$ with $0<\rk(\cF)<\rk(V)$.

Within the above set-up we will make a number of standard model-building choices. We are mostly interested in the observable sector and for the associated vector bundle $V$ we require a structure group $G={\rm SU}(n)$, where $n=3,4,5$. This means that the rank of $V$ should be ${\rm rk}(V)=n=3,4,5$ and, in order to have a special unitary rather than just a unitary structure group, we need
\begin{equation}
c_1(V)=0\; . \label{c1=0}
\end{equation} 
This class of bundles also enjoy nice properties with regard to stability. Examining \eref{slope}, we see that an $SU(n)$ bundle is stable if and only if all its proper sub-sheafs have {\em strictly} negative slope.
An immediate consequence of stability of $V$ is that $H^0(X,V)$ vanishes
\footnote{This ensues from the following simple argument. Since if $H^0(X,V) = H^0(X, V \otimes \cO_X^\star ) = \hom_X(\cO,V) \ne 0$, then $\cO_X$ is a proper sub-line-bundle of $V$. However, $c_1(\cO_X) = 0$ and $\rk(\cO_X)=1$ so $\mu(\cO_X) = 0$ and not strictly negative, whereby making $\cO_X$ a proper de-stablising subsheaf and $V$ would be unstable. Similarly, $H^0(X,V^\star ) = 0$.}:
Another useful property is that a bundle $V$ is stable if and only if its dual $V^\star$ is stable \cite{AG,monadbook}. Thus, $H^0(X,V^{\star})$ also vanishes for our bundles. In summary, for stable bundles $V$ we necessarily have
\beq\label{H0=0}
H^0(X,V) = H^0(X,V^\star ) = 0 \ .
\eeq
In fact \eref{H0=0} is only the first of a set of vanishing conditions (see Ref.~\cite{Donagi:2004ia}): if an $SU(n)$ bundle $V$ is stable, it is further true that
\beq\label{H0}
H^0(X, \wedge^p V) = 0 \qquad \forall~p = 1, \ldots, \rk(V)-1 \ .
\eeq
Note that since for $SU(n)$ bundles, $\wedge^p V \simeq \wedge^q V^\star$ for all $p+q=n=\rk(V)$  (cf.~Eq.\eref{SUn}), \eref{H0} is equivalent to saying that $H^0(X, \wedge^p V^{\star}) = 0$ for $p = 1, \ldots, \rk(V)-1$.

The definition of stability involves all coherent sub-sheafs of a given bundle  and is, therefore, typically not easy to prove. In this paper, we will be content with performing a ``check" for stability by verifying the necessary and highly non-trivial (but generally not sufficient) conditions \eref{H0}. 
A full stability proof of the monad bundles considered in this paper will appear in Ref.~\cite{stabpaper}.

Of course, we have to make sure that there exist a solution to the anomaly condition~\eqref{anomaly}. 
An effective way to guarantee this which does not require searching for suitable hidden bundles $\tilde{V}$ is to demand that
\begin{equation}
 c_2(TX)-c_2(V) \mbox{ is an effective class on } X\; . \label{effcond}
\end{equation}
In this case, both the anomaly and effectiveness conditions are satisfied~\footnote{Of course, there may be other choices  which involve a non-trivial hidden bundle $\tilde{V}$. Since we are mostly interested in the observable sector at this stage the important point for now is the existence of a viable hidden sector.} for a trivial hidden bundle $\tilde{V}$ and a five-brane class $W=c_2(TX)-c_2(V)$. 

The observable low-energy particle content from such a model is summarized in Table~\ref{t:spec}. For the three choices of structure group $G={\rm SU}(3), {\rm SU}(4)$ or  ${\rm SU}(5)$ one obtains low-energy GUTs with gauge group $H=E_6$,  ${\rm SO}(10)$ or ${\rm SU}(5)$, respectively.  The representations of $H$ which arise in the effective four-dimensional theory are obtained by decomposing the ${\bf 248}$ adjoint representation of $E_8$ under $G\times H$. The number of matter fields in the various representations is given by the dimension $h^1(X,U)$ of the bundle cohomology groups, where $U = V, V^\star , \wedge^2 V, \wedge^2 V^\star , V \otimes V^\star $, as indicated in Table~\ref{t:spec}.  A particularly useful quantity is the index ${\rm ind}(V)\equiv h^0(X,V)-h^1(X,V)+h^2(X,V)-h^3(X,V)$ of a bundle $V$. A stable bundle $V$ satisfies Eq.~\eqref{H0=0} and, hence, the index  equals ${\rm ind}(V)=-h^1(X,V)+h^2(X,V)$. Comparing with Table~\ref{t:spec}, this is precisely the chiral asymmetry, that is the difference of the number of anti-families and families. From the Atiyah-Singer index theorem it can be computed in terms of the third Chern class $c_3(V)$ of the bundle $V$ as
\begin{equation}
\ind(V)=-h^1(X,V)+h^2(X,V)=\frac{1}{2}\int_{X}c_{3}(V)\; . \label{indtheorem}
\end{equation}
Provided that $H^0(X,\Lambda^2 V)=H^0(X,\Lambda^2 V^\star)=0$, as will indeed be the case for our bundles and will be explicitly checked later on, the index theorem applied to $\Lambda^2 V$ together with the relation $c_3(\Lambda^2 V)=(n-4)c_3(V)$ (see the appendix of Ref.~\cite{Donagi:2004ia}) leads to 
\beq
(n-4) \ind(V) = - h^{1}(X,\wedge^2V) + h^{1}(X,\wedge^2 V^\star ) \ . \label{indtheorem2}
\eeq
This result will be useful for the ${\rm SU}(5)$ case and it implies that the chiral asymmetry between ${\bf\bar{5}}$ and ${\bf5}$ representations is the same as the one between ${\bf 10}$ and ${\bf\bar{10}}$. Hence, the chiral part of the spectrum always comes in pairs of ${\bf 10}$ and $\overline{\bf 5}$ (or $\overline{\bf 10}$ and ${\bf 5}$), that is in complete ${\rm SU}(5)$ families (or anti-families).
\begin{table}[h]
\begin{center}
\begin{tabular}{|l|l|l|}
\hline
$G\times H$ & Breaking Pattern:  
${\bf 248}\rightarrow $ 
& Particle Spectrum\\  
\hline\hline
{\small $\rm{SU}(3)\times E_{6}$} & {\small $({\bf 1},{\bf
  78})\oplus ({\bf 3},{\bf 27})\oplus 
(\overline{\bf 3},\overline{\bf 27})\oplus ({\bf 8},{\bf 1})$ }
&
$
\ba{rcl}
n_{27}&=&h^{1}(X,V)\\ 
n_{\overline{27}}&=&h^{1}(X,V^\star)=h^{2}(X,V)\\
n_{1}&=&h^{1}(X,V\otimes V^\star)
\ea
$
\\  \hline
{\small $\rm{SU}(4)\times\rm{SO}(10)$} &{\small $({\bf 1},{\bf
  45})\oplus ({\bf 4},{\bf 16}) 
\oplus (\overline{\bf 4},\overline{\bf 16})\oplus ({\bf 6},{\bf
  10})\oplus ({\bf 15},{\bf 1})$ } 
&
$
\ba{rcl}
n_{16}&=&h^{1}(X,V)\\
n_{\overline{16}}&=&h^{1}(X,V^\star)=h^2(X,V)\\
n_{10}&=&h^{1}(X,\wedge ^{2}V)\\
n_{1}&=&h^{1}(X,V\otimes V^\star)
\ea
$
\\  \hline
{\small $\rm{SU}(5)\times\rm{SU}(5)$} &{\small $({\bf 1},{\bf 24})\oplus
({\bf 5},{\bf 10})\oplus (\overline{\bf 5},\overline{\bf 10})\oplus
({\bf 10},\overline{\bf 5})\oplus 
(\overline{\bf 10},{\bf 5})\oplus ({\bf 24},{\bf 1})$}
&
$
\ba{rcl}
n_{10}&=&h^{1}(X,V)\\ 
n_{\overline{10}}&=&h^{1}(X,V^\star)=h^2(V)\\ 
n_{5}&=&h^{1}(X,\wedge^{2}V^\star)\\
n_{\overline{5}}&=&h^{1}(X,\wedge ^{2}V)\\
n_{1}&=&h^{1}(X,V\otimes V^\star)
\ea
$
\\ \hline
\end{tabular}
\caption{{\sf A vector bundle $V$ with structure group $G$ can break the $E_8$
  gauge group of the heterotic string into a GUT group $H$. The low-energy representation are found from the branching of the ${\bf 248}$ adjoint of $E_8$ under $G\times H$ and the low-energy spectrum is obtained by computing the indicated bundle cohomology groups.}}\label{t:spec}
\end{center}
\end{table}

For a realistic model the GUT group $H$ will have to be eventually broken to the standard model group. This is usually accomplished by dividing the Calabi-Yau manifold $X$ by a freely-acting discrete symmetry $\Gamma$ and then introducing Wilson lines on the quotient space $X/\Gamma$. In this paper, we will not carry this step out explicitly. However, when we analyse the properties of our models later on we will impose an important physical constraint which follows from this construction. Assuming that the bundle $V$ descends to the quotient space $X/\Gamma$ the ``downstairs" chiral asymmetry is given by ${\rm ind}(V)/|\Gamma |$, where $|\Gamma |$ is the order of the discrete symmetry group. Clearly, the downstairs chiral asymmetry should be three, so we should require that
\begin{equation}
 {\rm ind}(V) \mbox{ is divisible by } 3\; . \label{3gen}
\end{equation} 
Assuming this is the case, the order of the discrete group $\Gamma$ one needs is given by $|\Gamma |={\rm ind}(V)/3$. The Calabi-Yau manifold $X$ can only be quotiented by such a group if the Euler number $\chi (X)$ is divisible by its order $|\Gamma |$. Hence, in addition we demand that
\begin{equation}
 \chi (X) \mbox{ is divisible by } {\rm ind}(V)/3 \label{3gen1}
\end{equation} 
These two conditions are clearly necessary for successful Wilson line breaking to a model with three families but by no means sufficient. Nevertheless, we will see that they already impose strong constraints on monad bundles.

%
%
\section{Complete Intersection Calabi-Yau Threefolds}\setall
To begin our construction of vector bundles for heterotic models, we first discuss the relevant class of compact Calabi-Yau manifolds. Ever since the realization that Calabi-Yau three-folds played a central role in superstring compactification \cite{Candelas:1985en}, constructions of so-called ``complete intersection Calabi-Yaus'' (CICYs) \cite{Candelas:1987kf,Candelas:1987du,Green:1987cr,Green:1988,He:1990pg,Gagnon:1994ek} 
have been a topic of interest. Indeed, this method of Calabi-Yau construction was used in one of the first attempts to systematically study families of Calabi-Yau manifolds. Subsequent work, especially in
light of mirror symmetry, was carried out in explicit mathematical
detail \cite{Green:1987cr,He:1990pg,Gagnon:1994ek,Hosono:1994ax} for
half a decade, culminating in the pedagogical text \cite{hubsch} 
on the subject. The manifolds in \cite{Anderson:2007nc}, used to
illustrate a new algorithmic approach in heterotic compactification,
are special cases of these CICYs.

Unfortunately, much of the original data was stored on computer media,
such as magnetic tapes at CERN, which have
been rendered obsolete by progress. Partial results,
including, luckily, the list of the CICY threefolds itself, can be found on
the Calabi-Yau Homepage \cite{cypage}. In this section, we shall
resurrect some of the useful facts concerning the CICY threefolds,
which will be of importance to our bundle constructions later.
We will present only the essentials, leaving most of the details to Appendix
\ref{a:cicy}.

\subsection{Configuration Matrices and Classification}
We are interested in manifolds $X$ which can be described as algebraic varieties, that is, as intersections of the zero loci of $K$ polynomials $\{p_j\}_{j=1,\ldots ,K}$ in an ambient space $\cA$. For our purpose, we will consider ambient spaces $\cA =\IP^{n_1} \times \ldots \times \IP^{n_m}$ given by a product of $m$ ordinary projective spaces with dimensions $n_r$. We denote the projective coordinates of each factor $\IP^{n_r}$ by 
${\bf x}^{(r)} = [x_0^{(r)}:x_1^{(r)}:\ldots:x_{n_r}^{(r)}]$, its K\"ahler form by $J_r$ and the $k^{\rm th}$ power of the hyperplane bundle by $\cO_{\IP^{n_r}}(k)$. The K\"ahler forms are normalised such that
\begin{equation}
 \int_{P^{n_r}}J_r^{n_r}=1\; .
\end{equation}
The manifold $X$ is called a {\it complete intersection} if the dimension of $X$ is equal the dimension of $\cA$ minus the number of polynomials. This is, in a sense, the optimal way in which polynomials can intersect. To obtain threefolds $X$  from complete intersections we then need
\beq\label{ci}
\sum_{r=1}^m n_r - K = 3 \ .
\eeq
Each of the defining homogeneous polynomials $p_j$ can be characterised by its multi-degree ${\bf q}_j=(q_j^1,\ldots , q_j^m)$, where $q_j^r$ specifies the degree of $p_j$ in the coordinates ${\bf x}^{(r)}$ of the factor $\IP^{n_r}$ in $\cA$.  A convenient way to encode this information is by a {\it configuration matrix}
\beq\label{cy-config}
\left[\ba{c|cccc}
\IP^{n_1} & q_{1}^{1} & q_{2}^{1} & \ldots & q_{K}^{1} \\
\IP^{n_2} & q_{1}^{2} & q_{2}^{2} & \ldots & q_{K}^{2} \\
\vdots & \vdots & \vdots & \ddots & \vdots \\
\IP^{n_m} & q_{1}^{m} & q_{2}^{m} & \ldots & q_{K}^{m} \\
\ea\right]_{m \times K}\; .
\eeq
Note that the $j^{\rm th}$ column of this matrix contains the multi-degree of the polynomial $p_j$.
In order that the resulting manifold be Calabi-Yau, the condition
\beq\label{cy-deg}
\sum_{j=1}^K q^{r}_{j} = n_r + 1 \qquad \forall r=1, \ldots, m
\eeq
needs to imposed (essentially to guarantee that $c_1(TX)$ vanishes).
Henceforth, a CICY shall mean a Calabi-Yau threefold, specified by the
configuration matrix \eref{cy-config}, satisfying the conditions
\eref{ci} and \eref{cy-deg}. In fact, the condition \eref{cy-deg} even
obviates the need for the first column $\IP^{n_1} \ldots \IP^{n_m}$ in the
configuration matrix. Subsequently, we will frequently need the normal bundle $\cN$ of $X$ in $\cA$ which is given by
\begin{equation}
 \cN = \bigoplus_{j=1}^K\cO_\cA ({\bf q}_j)\; . \label{normalbundle}
\end{equation}
Here and in the following we employ the short-hand notation $\cO_\cA ({\bf k})=\cO_{\IP^{n_1}}(k^1)\otimes\dots\otimes\cO_{\IP^{n_r}}(k^r)$ for line bundles on the ambient space $\cA$. 

As an archetypal example, the famous quintic in $\IP^4$ is simply
denoted as ``$[4 | 5]$'', or, even more succinctly, as ``$[5]$''. One might
immediately ask how many possible non-isomorphic (one obvious
isomorphism being row and column permutations) configurations could
there be. This question was nicely settled in
\cite{Candelas:1987kf,He:1990pg} and the number is, remarkably,
finite. A total of 7890 is found and can be accessed at \cite{cypage}. This was the first large
data-set of Calabi-Yau manifolds (cf.~\cite{Candelas:2007ac}).

We have compiled an electronic list of these CICYs which contains all the essential information including configuration matrices, Euler numbers 
$\chi (X)$, second Chern classes $c_2(TX)$, Hodge numbers $h^{1,1}(X)$ and $h^{2,1}(X)$ and allows for easy calculation of triple intersection numbers. It also contains previously unknown information, in particular about redundancies within the CICY list. This data underlies many of the subsequent calculations for monad bundles on CICYs. For more details on this ``legacy" subject see Appendix~\ref{a:cicy}.

%
\subsection{Favourable Configurations}
Our choice of complete intersection Calabi-Yau manifolds is motivated largely by the explicit and relatively simple nature of the constructions. Perhaps the most valuable advantage of the presence of the ambient space $\cA$ is the existence of relatively straightforward methods to identify discrete symmetries, a crucial step for the implementation of Wilson line breaking. To take maximal advantage of the presence of the ambient space we will focus on CICYs for which this explicit embedding is particularly useful. For some CICYs, the second cohomology $H^2(X)$ is not entirely spanned by the restrictions of the ambient space K\"ahler forms $J_r$. For example, in the case of
the well-known Tian-Yau manifold, $X=\left[
\begin{array}
[c]{c}%
3\\
3
\end{array}
\left|
\begin{array}
[c]{ccc}%
3 & 0 & 1\\
0 & 3 & 1
\end{array}
\right.  \right]  $, there are two K\"ahler forms descending from the two $\mathbb{P}^{3}$'s, but $h^{1,1}(X)=14$. Here, we would like to focus on CICYs $X$ for which the second cohomology is entirely spanned by the ambient space K\"ahler forms and which are, hence, characterised by
\[
h^{1,1}(X)=m=\#\text{ of }\mathbb{P}^{n}\text{'s.}
\]
We shall call manifolds with this property \emph{favourable}. Such favourable CICYs offer a number of considerable practical advantages. There are 5 manifolds with $h^{1,1}(X)=m=1$. These are also referred to as {\em cyclic} CICYs and they constitute the subject of Ref.~\cite{Anderson:2007nc}.

The K\"ahler cone, that is the set of allowed K\"ahler forms $J$ on $X$, is simply given by $\{J=t^rJ_r\,|\,t^r\geq 0\}$, where $t^r$ are the K\"ahler moduli. Further, the set of all line bundles on $X$, the Picard group $\mbox{Pic}(X)$, is isomorphic to $\mathbb{Z}^{m}$, so  line bundles on $X$ can be characterised by an integer vector ${\bf k}=(k^1,\ldots ,k^m)$. We denote these line bundles by ${\cal O}_X({\bf k})$ and they can be obtained by restricting their ambient space counterparts $\cO_\cA ({\bf k})$ to $X$. 

We can also introduce a dual basis $\{\nu^r \}$ of $H^4(X,\mathbb{Z})$, satisfying
\begin{equation}
 \int_X\nu_r\wedge J_s=\delta^r_s\; , \label{nur}
\end{equation} 
and, via Poincar\'e duality $H^4(X,\mathbb{Z})\simeq H_2(X,\mathbb{Z})$, we can use this basis to describe the second integer homology of $X$. The effective classes $W\in H_2(X,\mathbb{Z})$ then correspond precisely to the  positive integer linear combinations of $\nu^r$, that is $w_r\nu^r$ with $w_r\geq 0$. This property makes checking our version of the anomaly cancellation condition~\eqref{effcond} very simple. If we expand second Chern classes in the basis $\{\nu^r \}$, writing $c_2(U)=c_{2r}(U)\nu^r$ for any bundle $U$, then the condition~\eqref{effcond} simply turns into the inequalities
\begin{equation}
 c_{2r}(V)\leq c_{2r}(TX) \quad \forall ~r = 1, \ldots, m. \label{effcond1}
\end{equation} 
Details on the computation of Chern classes on CICYs are given in Appendix~\ref{a:cicy}.

Scanning through the CICY data, we find that  there is a total of 4515 CICYs which are favourable. This is still a large dataset and we shall henceforth restrict our attention to these.

\subsection{Line Bundles on CICYs}
As we will see line bundles on CICYs are the main building blocks of the monad bundles considered in this paper, so we need to know their detailed properties. In particular we need to be able to fully determine the cohomology of line bundles on CICYs. We will return to this problem shortly after briefly reviewing a few more elementary properties. For an ambient space $\cA$ with $m$ projective factors, we consider a generic line bundle $L=\cO_X({\bf k})$ on a CICY $X$, where ${\bf k}=(k^1,\ldots ,k^m)$ is an $m$-dimensional integer vector. The Chern characters of such a line bundle are given by 
\beq\ba{rl}
\ch_{1}(L) & = c_1(L) = k^rJ_r\\
\ch_{2}(L) & =\frac{1}{2}k^rk^sJ_r\wedge J_s\\
\ch_{3}(L) & =\frac{1}{6}k^rk^sk^tJ_r\wedge J_s\wedge J_t \ ,
\ea\eeq
with implicit summation in $r,s,t = 1, \ldots, m$. Note that every line bundle on a CY $3$-fold is uniquely classified by its first Chern class, as can be seen explicitly from the above expression for $\ch_1$. The dual of the line bundle $L$ is simply given by $L^\star=\cO_X(-{\bf k})$. Using the Atiyah-Singer index theorem, the index of $L$ can be written as
\begin{align}\label{ind}
\ind(L)  &\equiv\sum_{q=0}^{3}(-1)^{q}h^{q}(X,L) = \int_X
\ch(L)\wedge\td(X)
=\int_{X}\left[  \ch_{3}(L)+\frac{1}{12}\ch_{2}(TX)\wedge
c_{1}(L)\right]  \nonumber\\
& =\frac{1}{6}\left( d_{rst} k^rk^sk^t+\frac{1}{2}k^rc_{2r}
(TX)\right)\; .
\end{align}

A special class of line bundles are the so-called {\em positive line bundles} which, in the present case, are the line bundles $L=\cO_X({\bf k})$ with all $k^r>0$.  The Kodaira vanishing theorem \eref{kodaira} applies to such positive bundles and (given the canonical bundle $K_X$ of a Calabi-Yau manifold is trivial) it implies that $h^{q}(X,L)=0$ for all $q\neq 0$. This means that  $h^{0}(X,L)$ is the only non-vanishing cohomology and it can, hence, be easily calculated from the index \eref{ind} since $h^0(X,L) = \ind(L)$. The situation is just as simple for {\em negative line bundles} $L$, that is line bundles for which $L^\star$ is positive. In our case, the negative line bundles $L=\cO_X({\bf k})$ are of course the ones with all $k^r<0$. Applying the Kodaira vanishing theorem to $L^\star =\cO (-{\bf k})$ and then using Serre duality it follows that $h^3(X,L)$ is the only non-vanishing cohomology of a negative line bundle. Again, it can be computed from the index using $h^3(X,L)=-{\rm ind}(V)$. These result for positive and negative line bundles can also be checked using the techniques of spectral sequences. In this case, the dimension of the single non-zero cohomology can be computed without explicitly knowledge of the Leray maps $d_{i}$  between cohomologies.  

One more general statement can be made. It turns out that semi-positive line bundles, that is line bundles  $L=\cO_X({\bf k})$, where $k^r\geq 0$ for all $r$, always have at least one section, so $h^0(X,L)>0$. One might be tempted to conclude that the line bundles with sections are precisely the semi-positive ones. While this is indeed the case for some CICYs it is by no means always true and for some CICYs the class of line bundles with a section is genuinely larger than the class of semi-positive line bundles.

Further quantitative statements about the cohomology of line bundles $L=\cO ({\bf k})$ containing ``mixed" or zero entries $k^{r}$ are not so easily obtained. For a general line bundle with mixed sign or zero entries, computing the dimensions $h^{q}(X,\mathcal{O}_{X}({\bf k}))$ does require explicit information about  the dimensions of kernels and ranks of the Leray maps $d_i$. Fortunately, this information can be obtained based on a computational variation of the Bott-Borel-Weil theorem. In this way, we are able to calculate all line bundle cohomologies on favourable CICYs explicitly.  We do not
particularly require this computation in the present paper and will defer a full discussion on the
matter to Ref.~\cite{stabpaper}.
The general result involves a large number of case distinctions, analogous to but significantly more complex than the Bott formula \eref{bott} for line bundle cohomology over $\IP^n$.

As an illustration, we provide a ``generalised Bott formula" for mixed line bundles of the form $\cO_X(-k,m)$ with $k \geq 1$, and $m \geq 0$ on the manifold $X=\left[
\begin{array}
[c]{c}%
1\\
3
\end{array}
\left|
\begin{array}
[c]{ccc}%
2 \\
4
\end{array}
\right.  \right]  $. We find that
\beq\label{example_coho}
h^{q}(X, \mathcal{O}_{X}(-k,m))=\left\{
\begin{array}
[c]{ll}%
(k+1)\binom{m}{3}-(k-1)\binom{m+3}{3} & q=0\quad k<\frac{(1+2m)(6+m+m^2)}{3(2+3m(1-m))}\\
(k-1)\binom{m+3}{3}- (k+1)\binom{m}{3} & q=1\quad k>\frac{(1+2m)(6+m+m^2)}{3(2+3m(1-m))}\\
0 & \mbox{otherwise}
\end{array}
\right. \ .
\eeq
where $\binom{n}{m}$ is the usual binomial coefficient with the convention that $\binom{0}{m}=1$.

It should be clear from the above example that the explicit formulae for mixed line bundle cohomology are complicated and, in practice, have to be implemented as a computer program. The outline of our algorithm for computer implementation will be presented in \cite{stabpaper}.

\section{The Monad Construction on CICYs}
As was discussed in Ref.~\cite{Anderson:2007nc},
large classes of vector bundles can be
constructed over projective varieties using a variant of Horrock's monad construction \cite{HM,beilinson,maruyama}. 
The monad-bundles have been used extensively in string compactification throughout the years 
\cite{Distler:1987ee,Blumenhagen:2006wj,Kachru:1995em,Blumenhagen:1997vt,Douglas:2004yv,maria}.
Vector bundles defined through the monad short exact sequences can be thought
of as kernels of maps between direct sums of line bundles. For reviews of this construction and some of its applications, see Ref.~\cite{monadbook,huybrechts,Blumenhagen:1997vt}.  The {\em monad bundles} $V$ considered in this paper are defined through the short exact sequence
\bea \label{monad}
\nn &&0 \to V \to B \stackrel{f}{\longrightarrow} C \to 0\ ,
\mbox{ where} \\
B &=& \bigoplus_{i=1}^{r_B} \cO_X({\bf b}_i) \ , \quad
C = \bigoplus_{j=1}^{r_C} \cO_X({\bf c}_j)
\label{defV}
\eea
are sums of line bundles with ranks $r_B$ and $r_C$, respectively. From the exactness of \eref{defV}, it follows that the bundle $V$ is defined as
\beq
V=\ker(f) \ .
\eeq
The rank $n$ of $V$ is easily seen, by exactness of \eref{defV}, to be
\beq
n = \rk(V) = r_B - r_C \ .
\eeq
Because the Calabi-Yau manifolds discussed in this work are defined as
complete intersection hypersurfaces in a product of projective spaces,
we can write a short exact sequence analogous to \eref{defV} but over the
ambient space, $\cA$.
\bea\label{ambmonad}
\nn &&0 \to \cV \to \cB \stackrel{\tilde{f}} {\longrightarrow} \cC \to 0\ ,\mbox{ where} \\
\cB &=& \bigoplus_{i=1}^{r_B} \cO_\cA({\bf b}_i) \ , \quad
\cC = \bigoplus_{j=1}^{r_C} \cO_\cA({\bf c}_j) \ .
\eea
Here, the map $\tilde{f}$ is a matrix whose entries are homogeneous polynomials of (multi-)degree ${\bf c}_j -{\bf b}_i$. The sequence \eref{ambmonad} defines a coherent sheaf $\cV$ on $\cA$ whose restriction to $X$ is $V$ (and additionally the map $f$ can be viewed as the restriction of $\tilde{f}$).\\

\noindent A number of mathematical constraints should be imposed on the above monad construction.
\paragraph{Bundleness: } It is not a priori obvious that the exact sequence~\eqref{monad} indeed defines a bundle rather than a coherent sheaf. However, thanks to a theorem of Fulton and Lazarsfeld~\cite{lazarsfeld} this is the case provided two conditions are satisfied 
(see also~\cite{Anderson:2007nc}). 
First, all line bundles in $C$ should be greater or equal than all line bundles in $B$. By this we mean that $c^r_j\geq b^r_i$ for all $r$, $i$ and $j$. Second, the map $f:B\rightarrow C$ should be sufficiently generic~\footnote{The actual condition of Fulton and Lazarfeld's theorem, apart from genericity of $f$, is that $C^\star\otimes B$ is globally generated so has at least $r_B r_C$ sections. This is indeed the case if $c^r_j\geq b^r_i$ for all $r$, $i$ and $j$ since, in this case, the line bundles $\cO_X({\bf c}_j-{\bf b}_i)$ which make up $C^\star\otimes B$ are semi-positive so have at least one section each. On some CICYs the line bundles with sections extend beyond the semi-positive ones, as discussed earlier, and for those CICYs one can likely allow monads where some of the entries in $C$ are smaller than the ones in $B$ and still preserve ``bundleness" of $V$. In the present paper, we will not pursue this very case-dependent possibility further.}. Phrased in terms of ambient space language this means that the map $\tilde{f}:\cB\rightarrow\cC$ should be made up from sufficiently generic homogeneous polynomials  of degree ${\bf c}_j-{\bf b}_i$. We will henceforth require these two conditions. An immediate consequence of $V$ being a bundle is that \eqref{monad} can be dualized to the short exact sequence
\beq\label{dualV}
0 \to C^\star  \stackrel{f^T} {\longrightarrow} B^\star  \to V^\star  \to 0 \ ,
\eeq
so that the dual bundle $V^\star $ is given by
\beq
V^\star =\coker(f^T)\, .
\eeq

\paragraph{Non-triviliaty: } The above constraint on the integers $c^r_j$ and $b^r_i$ can be slightly strengthened. Suppose that a monad bundle $V$ is defined by the short exact sequence
\beq
0 \to V \to B \oplus R \stackrel{f'}{\longrightarrow} C \oplus R \to 0 \ ,
\eeq
where the repeated summand $R$ is a line bundle or direct sum of line bundles. The so-defined bundle $V$ is indeed equivalent to the one defined by the sequence~\eqref{monad}, so the common summand $R$ is, in fact, irrelevant\footnote{This follows directly from the Snake Lemma \cite{AG}, using the obvious injections of $B,C$ into $B\oplus R$ and $C\oplus R$.}. To exclude common line bundles in $B$ and $C$ we should demand that all line bundles in $C$ are strictly greater than all line bundles in $B$. By this we mean that $c^r_j\geq b^r_i$ for all $r$, $i$ and $j$ and, in addition, that for all $i$ and $j$ strict inequality, $c^r_j>b^r_i$, holds for at least one $r$ (which can depend on $i$ and $j$).

\paragraph{Positivity: } We require that all line bundles in $B$ and $C$ are positive, that is $b^r_i>0$ and $c^r_j>0$ for all $i$, $j$ and $r$. Monads discussed in the physics literature~\cite{Distler:1987ee,Blumenhagen:2006wj,Kachru:1995em,Blumenhagen:1997vt,Douglas:2004yv} 
have typically been of this type and we will refer to them as {\em positive monads}. The reasons for this constraint are mainly of a practical nature. We have seen in our discussion of line bundles on CICYs that the cohomology of positive line bundles is particularly simple and easy to calculate from the index theorem. 
This fact significantly simplifies the analysis of positive monads. 

Furthermore, experience seems to indicate that non-positive bundles are ``more likely" to be unstable. As an extreme case, one can easily show 
that monads constructed only from negative line bundles are unstable because they explicitly have non-vanishing $H^0(X, V^\star)$. 
Of course we are not implying that all non-positive monads are unstable. In fact, in a forthcoming paper~\cite{zeropaper} we will show that allowing zero entries can still be consistent with stability. However, from the point of view of stability, starting with positive monads seems the safest bet, and we will focus on this class in the present paper. 

In addition to the constraints of a more mathematical nature above we should consider physical constraints. To formulate them we need explicit expressions for the Chern classes of monad bundles. One finds
\bea
\nn \rk(V) &=& r_B - r_C = n  \ , \\
\nn c_1^r(V) &=& \sum_{i=1}^{r_B} b^r_i - \sum_{j=1}^{r_C} c^r_j \ ,
\\
c_{2r}(V &=& \frac12  d_{rst} 
   \left(\sum_{j=1}^{r_C} c^s_j c^t_j- 
   \sum_{i=1}^{r_B} b^s_i b^t_i \right) \ , 
\label{chernV} \\
\nn c_3(V) &=& \frac13 d_{rst} 
   \left(\sum_{i=1}^{r_B} b^r_i b^s_i b^t_i - \sum_{j=1}^{r_C} c^r_j
   c^s_j c^t_j \right) \ ,
\eea
where $d_{rst}$ are the triple intersection numbers~\eqref{drst} on $X$ and the relations for $c_{2r}(V)$ and $c_3(V)$ have been simplified assuming that $c_1^r(V)=0$. Then we need to impose two physics constraints.
\paragraph{Correct structure group:} To have bundles with structure group ${\rm SU}(n)$ where $n=3,4,5$ we first of all need that $n=r_B-r_C=3,4,5$. In addition,  the first Chern class of $V$ needs to vanish which, from the second Eq.~\eqref{chernV}, can be expressed as
\beq
S^r := \sum_{i=1}^{r_C+n} b^r_i = \sum_{j=1}^{r_C} c^r_j  \qquad \forall r=1,
   \ldots, k \ .
\label{cons1}
\eeq
We have defined the quantities $S^r$ which represent the first Chern classes of $B$ and $C$ and will be useful for the classification of positive monads below.

\paragraph{Anomaly cancellation/effectiveness:} As we have seen this condition can be stated in the simple form~\eqref{effcond1}. Inserting the above expression for the second Chern class gives
\beq
d_{rst} \left( \sum_{j=1}^{r_C} c^s_j c^t_j - \sum_{i=1}^{r_B} b^s_i b^t_i \right) \leq 2c_{2r}(TX) \qquad \forall r \ . \label{effcond2}
\eeq

In addition, we should of course prove stability of positive monads, a task which will be systematically dealt with in Ref.~\cite{stabpaper}. This completes the set-up of monads bundles. To summarise, we will consider monad bundles $V$ of rank $3$, $4$ or $5$, defined by the short exact sequence~\eqref{monad} with positive line bundles only. In addition, all line bundles in $C$ must be strictly greater than all line bundles in $B$ and the two constraints~\eqref{cons1} and \eqref{effcond2} must be satisfied.

%
\section{Classification of Positive Monads on CICYs}\setall\label{s:positive}
An obvious question is whether the class of monads defined in the previous section is finite. In this section, we show that this is indeed the case and subsequently classify all such monads.

We start by stating the classification problem in a more formal way. For any favourable CICY manifold $X$ with second Chern class $c_{2r}(TX)$ and triple intersection numbers $d_{rst}$, defined in a product of $m$ projective spaces, and for any $n=3,4,5$, we wish to find all sets of integers $b^r_i$ and $c^r_j$, where $r=1,\ldots ,m$, $i=1,\ldots ,r_B=r_C+n$ and $j=1,\ldots , r_C$ satisfying the conditions
\bea
\nn 1.&&b^r_i\geq 1\; ,\quad c^r_j\geq 1\; ,\quad \forall i,j,r;\\
\nn 2.&&c^r_j \ge b^r_j \ , \quad \forall i,j,r;\\ 
\nn 3. &&\forall i,j \mbox{ there exists at least one } r \mbox{ such that }c^r_j>b^r_i;\\
 4.&&\sum_{i=1}^{r_B} b^r_i = \sum_{j=1}^{r_C} c^r_j = S^r \ , \quad
    \forall r; \label{class}\\
\nn 5.&&d_{rst} \left( \sum_{j=1}^{r_C} c^s_j c^t_j - \sum_{i=1}^{r_B} b^s_i b^t_i \right) \leq 2c_{2r}(TX) \qquad \forall r \, .
\eea
Our first task is to show that this defines a finite class. Although all that is involved are simple manipulations of inequalities it is not complete obvious at first which approach to take. We start by defining the maxima $b^r_{\rm max}={\rm max}_i\{b^r_i\}$, minima $c^r_{\rm min}={\rm min}_j\{c^r_j\}$ and their differences $\theta^r=c^r_{\rm min}-b^r_{\rm max}\geq 0$ which are of course positive for all $r$. Then we can write
\beq
b^r_i = b^r_{\rm max} - T^r_i, \qquad
c^r_j = c^r_{\rm min} + D^r_j \ , \label{bcmaxmin}
\eeq
where $T^r_i$ and $D^r_j$ are the deviations from the maximum and minimum values. It is also useful to introduce the sums
\begin{equation}
 T^r=\sum_{i=1}^{r_B} T^r_i\; ,\quad D^r=\sum_{j=1}^{r_C} D^r_j \label{TDdef}
\end{equation}
of these deviations. Given theses definitions, it is easy to see that
\begin{equation}
S^r = b^r_{\rm max} r_B - T^r\; ,\quad S^r = c^r_{\rm min} r_C + D^r\; .
\end{equation}
Subtracting these two equations and using $r_B=r_C+n$ it follows that
\beq\label{thetaid}
\theta^rr_C + (D^r + T^r) = n b^r_{\rm max} \ .
\eeq
We will use this identity shortly.  Next, from the definition \eqref{cons1}, and since all $c_s^j \ge 1$, we
obtain the two inequalities
\beq\label{ineq2}
S^r \ge \sumj \II^r = r_C \II^r \; , \quad S^r \le \sumi b^r_{\rm max} = b^r_{\rm max} r_B\; ,\quad \forall~r \; ,
\eeq
where $\II_s$ is a vector with all entries being 1. After this preparation, we come to the key part of the argument which involves working out the consequences of condition 5 in \eqref{class}.
\beq\label{boundS}
\ba{llll}
2c_{2r}(TX) 
&\ge& 
d_{rst} \left(
\sumj c^s_j c^t_j - \sumi b^s_i b^t_i 
\right) & \\
&=&
d_{rst}\left(
\sumj (c^s_{\rm min}+D^s_j) c^t_j - 
\sumi (b^s_{\rm max}-T^s_i) b^t_i 
\right) & \mbox{inserting }\eqref{bcmaxmin}\\
&=& d_{rst}\left(
(c^s_{\rm min} - b^s_{\rm max}) S^t + \sumj D^s_j c^t_j + 
\sumi T^s_i b^t_i 
\right) & \mbox{using }\eqref{cons1}\\
&\ge& d_{rst}\left(
\theta^s S^t + (D^s + T^s) \II_t
\right) & \mbox{since } c^t_j, b^t_i \ge 1\ ,\mbox{using }\eqref{TDdef}\\
&\ge& d_{rst}\left(
\theta^s (r_C \II^t) + (D^s + T^s) \II^t 
\right) & \mbox{by first inequality }\eref{ineq2}\\
&=& d_{rst}\left(
n b^s_{\rm max} \II^t
\right) & \mbox{from \eref{thetaid}}\\
&\ge& \frac{n}{r_B} 
d_{rst}\left( S^s \II^t
\right) & \mbox{by second inequality}
\ea
\eeq
From the second last line in the above chain of inequalities, we can also express this result as a bound in the variables $b^r_{\rm max}$ (the maximum entries the bundle $B$ can have in each projective space), resulting in
\beq\label{bsmaxineq}
2c_{2r}(TX) \ge n\sum_{s,t} d_{rst} b^s_{\rm max} \ .
\eeq
It turns out that the matrices $\sum_td_{rst}$ are always non-singular, so this equation provides an upper bound for $b^r_{\rm max}$. Moreover, since each $b^r_{\rm max} \in \IZ_{\ge 1}$, and since the
matrix $n \sum_t d_{rst}$ has entries in $\IZ_{\ge 0}$,
Eq.~\eref{bsmaxineq} may not have solutions for all manifolds. In fact, of
the 4515 favourable CICYs, Eq.~\eref{bsmaxineq} immediately 
eliminates all but 63 which include the 5 cyclic ones studied in Ref.~\cite{Anderson:2007nc}. 
One finds that the values for $b^r_{\rm max}$ are very small indeed and never exceed 6.
\comment{
{7643, 7668, 7707, 7708, 7725, 7727, 7728, 7758, 7759, 7779, 7789, 7797,
7799, 7806, 7807, 7808, 7809, 7816, 7817, 7819, 7821, 7822, 7823, 7831, 7833,
7834, 7836, 7840, 7844, 7845, 7853, 7854, 7858, 7859, 7861, 7862, 7863, 7865,
7866, 7867, 7868, 7869, 7870, 7871, 7872, 7873, 7874, 7875, 7876, 7877, 7878,
7879, 7880, 7881, 7882, 7883, 7884, 7885, 7886, 7887, 7888, 7889, 7890}
}

So far, we have bounded the maximal entries of the bundle $B$. What about $r_B$, the rank of $B$? It turns out there are various ways to derive an upper bound on $r_B$.  First note that, from the third condition in \eref{class}, for all $j\in \{1,\ldots ,r_C\}$, there exists a $\sigma\in \{1,\ldots, m\}$, call it $\sigma (j)$, such that
\begin{equation}
c^r_j-b^r_{\rm max} \geq \delta^{r \sigma(j)} \; .
\end{equation}
Introduce
\begin{equation}\label{def-nu}
\nu^r = \sum_{j=1}^{r_C}\delta^{r \sigma(j)} \; ,
\end{equation}
the number of line bundles in $C$ which are bigger than the ones in $B$ due to the $r$-th entry. 
Since all line bundles in $C$ are bigger than the ones in $B$ it follows that
 \begin{equation}
 \sum_{r=1}^m \nu^r = r_C = r_B + n\; .\label{nusum}
 \end{equation}
We conclude that
 \begin{equation}
 r_Bb^r_{\rm max}\geq\sum_{i=1}^{r_B}b^r_i=\sum_{j=1}^{r_C}c^r_j
 \geq\sum_{j=1}^{r_C}(b^r_{\rm max}+\delta^{r\sigma (j)})=r_cb^r_{\rm max}+\nu^r
 \end{equation}
and, hence, that $nb^r_{\rm max}\geq\nu^r$. Summing this result over $r$ one easily finds that
 \begin{equation}
  r_B\leq n\left(1+\sum_{r=1}^mb^r_{\rm max}\right)\; . \label{rBbound1}
 \end{equation}
Since we have already bounded $b^r_{\rm max}$ (independently of $r_B$) this provides an upper bound for $r_B$. This shows that our class of bundles is indeed finite. While the above bound is simple, for the practical purpose of classifying all bundles it often turns out to be too weak, and requires computationally expensive scanning of monads with large $r_B$ and, hence, a large number of integer entries. Based on Eq.~\eqref{rBbound1} alone, a classification on a desktop machine is likely impossible. Fortunately, one can derive other constraints on $r_B$ which in many cases turn out to be stronger. Using $nb^r_{\rm max}\geq\nu^r$ in Eq.~\eqref{bsmaxineq} leads to
 \begin{equation}
  \sum_{r,s}d_{rst}\nu^t\leq 2c_{2r}(TX)\; . \label{rBbound2}
\end{equation}
For each CICY, one can find all integer solutions $(\nu^r)$ (subject to the constraint
$\nu^r\geq 0$, of course) to this equation and then calculate the maximal possible value for $r_B$ from Eq.~\eqref{nusum}. Finally, starting again from condition 5 of \eref{class} we find
\beq\ba{rcl}
 2c_{2r}(TX)
&\geq& 
d_{rst}\left[
\sum\limits_{j=1}^{r_C}c^s_jc^t_j-\sum\limits_{i=1}^{r_B}b^s_ib^t_i\right]\\
&\geq& 
d_{rst}\left[\sum\limits_{j=1}^{r_C}(b^s_{\rm max}+\delta^{s\sigma (j)})
(b^t_{\rm max}+ \delta^{t\sigma (j)})- 
\sum\limits_{i=1}^{r_B}b^s_i b^t_i\right]\\
&=&
d_{rst}\left[\sum\limits_{j=1}^{r_C}b^s_{\rm max}b^t_{\rm max} -
\sum\limits_{i=1}^{r_B}b^s_i b^t_i+ 2\nu^s b^t_{\rm max} +
\delta_{s}^t \nu^t\right]\\
&\geq
&d_{rst}\left[ -n b^s_{\rm max} b^t_{\rm max} + 2\nu^s b^t_{\rm max}
+\delta_{s}^t\nu^t\right]\; .
\ea\eeq
Rewriting this as an system of linear inequalities for $\nu^s$, we have that
\begin{equation}\label{rBbound3}
\sum\limits_s\left(2\sum\limits_td_{rst}b^t_{\rm max}+d_{rss}\right)\nu^s\leq 2c_{2r}(TX)+nd_{rst}b^s_{\rm max}b^t_{\rm max}\; .
\end{equation}
Again, this equation can be solved for all non-negative integers $\nu^r$ since
$b^{max}_s$ is bounded from \eref{bsmaxineq} and, subsequently, we can compute the
maximal $r_B$ from Eq.~\eqref{nusum}. In practice, we evaluate all three bounds~\eqref{rBbound1}, \eqref{rBbound2}, \eqref{rBbound3} for every CICY and use the minimum value obtained. In this way we find maximal values for $r_B$ ranging from $8$ to $22$ depending on the CICY. 

The explicit classification is now simply a matter of computer search. For each of the 63 CICYs with solutions to the inequality~\eqref{bsmaxineq} we scan over all allowed values of $n$, $r_B$ and over all values for $S^r$ subject to the last inequality in \eqref{boundS}. For each fixed set of these quantities we then generate all multi-partitions of entries $b^r_i$ and $c^r_j$ eliminating, of course, trivial redundancies due to permutations.
\comment{
Computing $r_B$ from the three constraints \eqref{cons1}, \eqref{cons2} and \eqref{cons3} and taking the minimal value I find for the
63 CICYs which pass the $b_{\rm max}$ test I find (in pairs (CICY\#,maximal rank of $B$)):
\begin{equation}
\begin{array}{llllllll}
(7643, 8)& (7668, 8)& ( 7707, 11)& (7708, 11)& (7725, 8)& (7727, 11)& (7728, 11)& (7758, 8)\\
( 7759, 8)& (7779, 12)& (7789, 12)& (7797, 11)& (7799, 8)& (7806, 14)& (7807, 8)& (7808, 8)\\
(7809, 8)& (7816, 11)& (7817, 11)& (7819, 11)& (7821, 8)& (7822, 11)& (7823, 11)& (7831, 12)\\
(7833, 10)& (7834, 11)& (7836, 11)& (7840, 14)& (7844, 9)& (7845, 12)& (7853, 9)& (7854, 11)\\
(7858, 14)& (7859, 14)& (7861, 7)& (7862,16)& (7863, 9)& (7865, 11)& (7866, 11)& (7867, 14)\\
(7868, 10)& (7869,14)& (7870, 13)& (7871, 11)& (7872, 12)& (7873, 14)& (7874, 14)& (7875,14)\\
(7876, 11)& (7877, 14)& (7878, 9)& (7879,8)& (7880, 16)& (7881, 14)& (7882, 16)& (7883, 10)\\
(7884, 12)& (7885, 22)& (7886, 22)& (7887, 20)& (7888, 22)& (7889, 9)& (7890, 12)&
\end{array}\nonumber
 \end{equation}
}
Upon performing this scan, we find that positive monad bundles only exist over 36 favourable CICYs (out of the 63 which passed the initial test). These 36 manifolds, together with the number of monad bundles over them, are listed in Table~\ref{t:pos-cicy}. 
\begin{table}
\begin{center}
\beq\nn
\ba{|c|c||c|c||c|c||c|c|}\hline
\mbox{Config} & \mbox{No.Bundles} & \mbox{Config} & \mbox{No.Bundles} &
\mbox{Config} & \mbox{No.Bundles} & \mbox{Config} & \mbox{No.Bundles}
\\ \hline \hline
\tconf{5} & (20, 14, 9)
& \tconf{3 & 3 } & (5, 3, 2)
& \tconf{4 & 2 } & (7, 5, 3)
& \tconf{3 & 2 & 2 } & (3, 2, 1)
\\ \hline
\tconf{2 & 2 & 2 & 2 \cr} & (2, 1, 0)
& \tconf{ 2 \cr 4 \cr  } & (611, 308, 56)
& \tconf{ 3 \cr 3 \cr  } & (62, 43, 14)
& \tconf{0 & 2 \cr 2 & 3 \cr} & (80, 12, 0)
\\ \hline
\tconf{0 & 2 \cr 3 & 2 \cr  } & (12, 5,0)
& {}^{(4)}\tconf{0 & 2 \cr 4 & 1 \cr  } & (126, 17, 0)
& \tconf{ 1 & 1 \cr 3 & 2 \cr  } & (15, 8, 0)
& \tconf{ 1 & 1 \cr 4 & 1 \cr  } & (153, 35, 19)
\\ \hline
\tconf{ 2 & 1 \cr 1 & 3 \cr  } & (3, 0, 0)
& \tconf{ 2 & 1 \cr 2 & 2 \cr  } & (5, 0, 0)
& \tconf{ 2 & 1 \cr 3 & 1 \cr  } & (13, 2, 0)
& {}^{(2)}\tconf{ 0 & 0 & 2 \cr 2 & 2 & 2 \cr  } & (5, 0, 0)
\\ \hline
{}^{(3)}\tconf{ 0 & 0 & 2 \cr 3 & 2 & 1 \cr  } & (5, 0, 0)
& {}^{(2)}\tconf{ 0 & 1 & 1 \cr 2 & 2 & 2 \cr  } & (5, 0, 0)
& \tconf{ 0 & 1 & 1 \cr 2 & 3 & 1 \cr  } & (12, 5, 0)
& \tconf{ 0 & 1 & 1 \cr 3 & 2 & 1 \cr  } & (8, 0, 0)
\\ \hline
{}^{(4)}\tconf{ 0 & 1 & 1 \cr 4 & 1 & 1 \cr  } & (126, 17, 0)
& \tconf{ 0 & 2 & 1 \cr 2 & 2 & 1 \cr  } & (2, 0, 0)
& \tconf{ 1 & 1 & 1 \cr 3 & 1 & 1 \cr  } & (2, 0, 0)
& \tconf{ 2 & 1 & 1 \cr 2 & 1 & 1 \cr  }  & (1, 0, 0)
\\ \hline
\tconf{0 & 0 & 1 & 1 \cr 2 & 2 & 2 & 1 \cr  } & (3, 0, 0)
& {}^{(3)}\tconf{0 & 0 & 1 & 1 \cr 3 & 2 & 1 & 1 \cr  } & (5, 0, 0)
& \tconf{ 2 \cr 2 \cr 3 \cr  } & (553, 232, 0)
& \tconf{ 0 & 2 \cr 1 & 2 \cr 1 & 2 \cr  } & (8, 0, 0)
\\ \hline
\tconf{ 1 & 1 \cr 0 & 2 \cr 1 & 3 \cr  } & (74, 0, 0)
& {}^{(1)}\tconf{ 1 & 1 \cr 0 & 2 \cr 2 & 2 \cr  } & (9, 0, 0)
& \tconf{ 1 & 1 \cr 1 & 1 \cr 1 & 3 \cr  } & (25, 0, 0)
& {}^{(1)}\tconf{ 1 & 1 \cr 1 & 1 \cr 2 & 2 \cr  } & (9, 0, 0)
\\ \hline
\tconf{ 1 & 1 \cr 1 & 2 \cr 0 & 3 \cr  } & (34, 0, 0)
& \tconf{ 1 & 1 \cr 2 & 1 \cr 2 & 1 \cr  } & (3, 0, 0)
& \tconf{1 & 1 & 0 \cr 1 & 0 & 1 \cr 3 & 1 & 1 \cr  } & (9, 0, 0)
& \tconf{2 \cr 2 \cr 2 \cr 2 \cr } & (3665, 625, 0)
\\
\hline
\ea
\eeq
\end{center}
\caption{{\sf The 36 manifolds which admit positive monads. The No.Bundles column next
to each manifold is a triple, corresponding to the respective numbers of ranks 3,4,
and 5 monads. Identical numbers in brackets to the left of a configuration matrices indicate equivalent configurations as identified in Appendix~\ref{a:cicy}.}}
\label{t:pos-cicy}
\end{table}

\begin{figure}[t]
\centerline{(a)\epsfxsize=3in\epsfbox{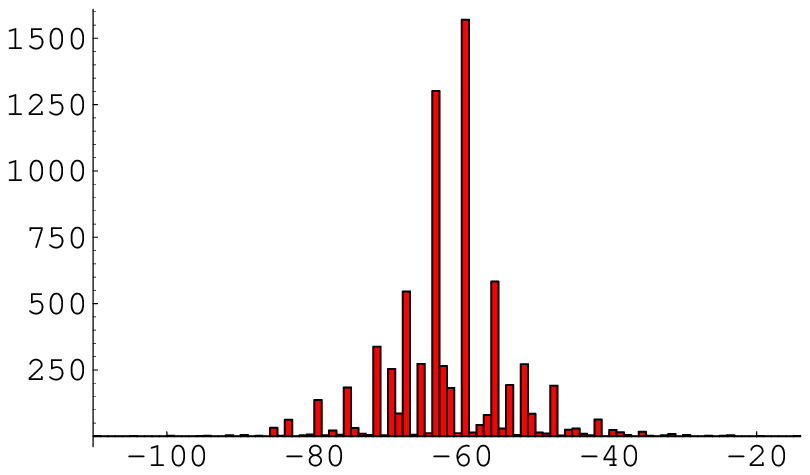}
(b)\epsfxsize=3in\epsfbox{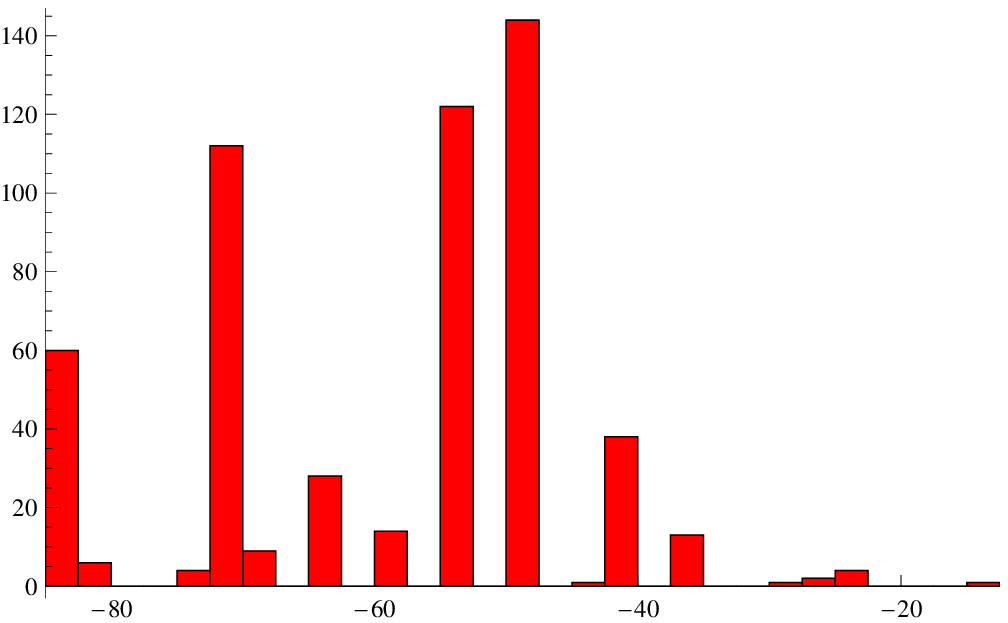}}
\caption{{\sf (a) Histogram for the index,
${\rm ind}(V)$, of the 7118 positive monads
found over 36 favourable CICYs: the horizontal axis is ${\rm ind}(V)$ and the
vertical, the number of bundles;
(b) the same data set,
but only taking into account monads with ${\rm ind}(V) = 3k$ for some positive 
integer $k$, such that $k$ divides the Euler number of the corresponding CICY.
}}
\label{f:c3pos}
\end{figure}

\begin{figure}[t]
\centerline{(a)\epsfxsize=3in\epsfbox{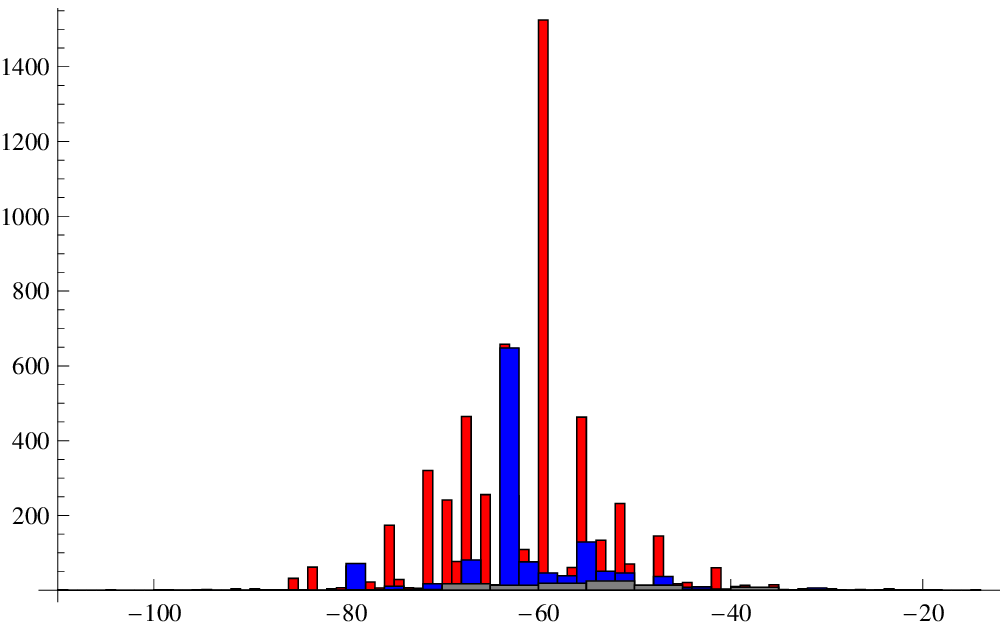}
(b)\epsfxsize=3in\epsfbox{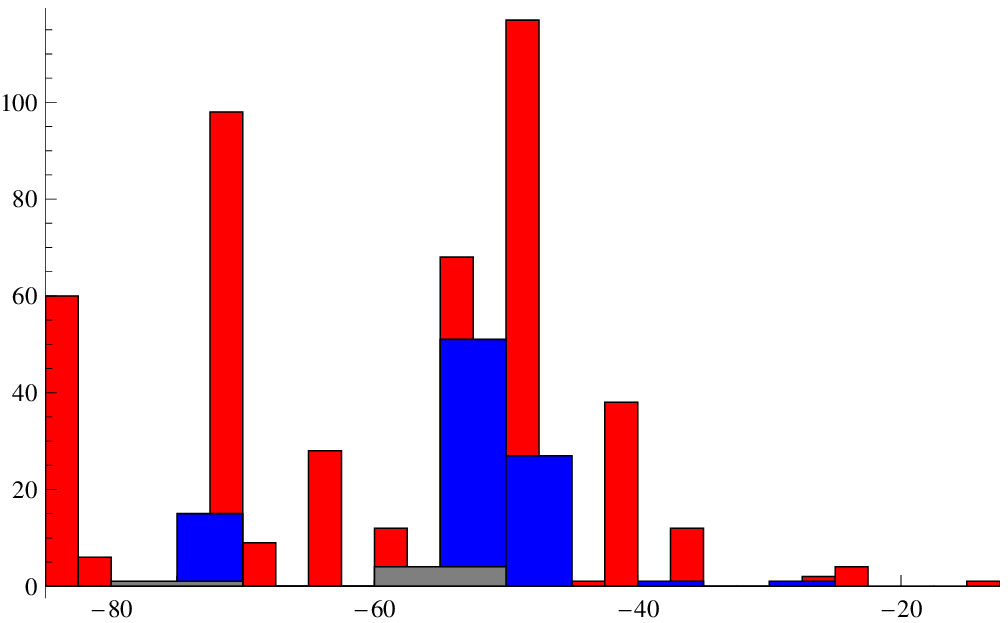}}
\caption{{\sf (a) Histogram for the index, ${\rm ind}(V)$, of the
positive monads, 5680 of rank 3 (in red), 1334 of rank 4 (in blue), and 104 of
rank 5 (in gray), found over 36 favourable CICYs: the horizontal axis is ${\rm ind}(V)$ and the
vertical, the number of bundles; (b) the same data set, but only taking into account monads with ${\rm ind}(V) = 3k$ for some positive integer $k$, such that $k$ divides the Euler number of the corresponding CICY.
}}
\label{f:c3pos345}
\end{figure}

In total, we find 7118 positive monad bundles. These include the 77 positive monad bundles on the 5 cyclic CICYs (these are the CICYs with $h^{1,1}(X)=1$) found in Ref.~\cite{Anderson:2007nc}.
Some explicit examples are listed in Table~\ref{t:monex}.
\begin{table}
\begin{center}
\beq\nn
\ba{|c|c|c|c|c|c|}\hline
\mbox{CICY } X & B & C & \rk(V) &
\left[ \ba{c} c_2(TX) \\ c_2(V) \ea \right] & \mbox{ind}(V) = \frac12c_3(V)
\\ \hline \hline
\left[\begin{array}{c|c}
1 & 2\\
1 & 2\\
1 & 2\\
1 & 2\\
\end{array}\right]
&
\cO_X(1,1,1,1)^{\oplus 8} &
\ba{c}
\cO_X(5, 1, 1, 1) \\
\oplus \cO_X(1, 5, 1, 1) \\
\oplus \cO_X(1, 1, 5, 1) \\
\oplus \cO_X(1, 1, 1, 5)
\ea
& 4 & \left[ \ba{c} (24, 24, 24, 24) \\ (24, 24, 24, 24) \ea \right]
& -64 \\ \hline
\left[\begin{array}{c|cc}
1 & 1 & 1\\
2 & 2 & 1\\
2 & 2 & 1\\
\end{array}\right]
& \cO_X(1,1,1)^{\oplus 10} &
\ba{c}
\cO_X(1, 1, 2)^{\oplus 3} \\
\oplus \cO_X(1, 2,1)^{\oplus 3} \\
\oplus \cO_X(4, 1, 1)
\ea
& 3 & \left[ \ba{c} (24, 36, 36) \\ (24, 36, 36) \ea \right]
& -69 \\ \hline
\left[\begin{array}{c|c}
1 & 2 \\
3 & 4 \\
\end{array}\right]
& \cO_X(1,1)^{\oplus 11}
& \cO_X(6,1) \oplus \cO_X(1,2)^{\oplus 5}
& 5 & \left[ \ba{c} (24, 44) \\ (20, 30) \ea \right]
& -40 \\ \hline
[4|5] & \cO_X(1)^{\oplus 6} & \cO_X(2)^{\oplus 3}
& 3 & \left[ \ba{c} (50) \\ (15) \ea \right]
& -15 \\ \hline
\ea\nn
\eeq
\caption{\sf Some examples from the 7118 positive monads on favourable CICYs.}
\label{t:monex}
\end{center}
\end{table}
Focusing on the different ranks of $V$ considered, we find 5680 bundles of rank 3, 1334 of rank 4, and 104 of
rank 5 on these 36 manifolds. To get an idea of the distribution, in part (a) of Fig.~\ref{f:c3pos} we have plotted the number of monads as a function of the index ${\rm ind}(V)$. It seems, at first glance, that the distribution is roughly Gaussian. For comparison, in part (b) of Fig.~\ref{f:c3pos}, we have plotted the number of monads which satisfy the two 3-generation constraints~\eqref{3gen} and \eqref{3gen1}. The same data, but split up into the three cases $n=3,4,5$ for the rank of $V$, is shown in Fig.~\ref{f:c3pos345}. The total numbers of bundles in all cases has been collected in Table~\ref{t:pos-monad}.
\begin{table}[h]
\begin{center}
\begin{tabular}{|c|c|c|c|c|}  \hline
  & Bundles & ${\rm ind}(V) = 3k$ & \begin{tabular}{l} ${\rm ind}(V) = 3k$ \\ and $k$ divides $\chi(X)$ \end{tabular} 
  & \begin{tabular}{l} ${\rm ind}(V) = 3k$ \\ $|{\rm ind}(V)|<40$ \\
  and $k$ divides $\chi(X)$ \end{tabular} \\ \hline \hline
rank 3 & 5680 & 3091 & 458 & 19\\
rank 4 & 1334 & 207 & 96 & 2 \\
rank 5 & 104 & 52 & 5 & 0 \\ \hline
Total & 7118 & 3350 & 559 & 21\\ 
\hline
\end{tabular}
\end{center}
\caption{{\sf The number of positive monad bundles on favourable CICYs.
Imposing that the index, ind$(V)$, is divisible by 3 reduces the number
and requiring, in addition, that ind$(V)/3$ divides the Euler number of the corresponding CICY leads to a further reduction.}}
\label{t:pos-monad}
\end{table}

It is clear from this table that even the two very rudimentary physical constraints~\eqref{3gen} and \eqref{3gen1} lead to a very substantial reduction of the number of viable bundles. If these two constraints are combined with a ``sensible" limit on the index, for example ${\rm ind}(V)< 40$ (assuming that the discrete symmetries one is likely to find are of order $\leq 13$), then part (b) of the figures show that the number of remaining bundles is very small indeed: there are only 21 of these. Remarkably these, perhaps of the most physical interest, only exist on the cyclic manifolds discussed in \cite{Anderson:2007nc} as well as the transposes of these configuration matrices (cf.~\cite{Candelas:2007ac}).

\section{Stability}\label{s:stable}
\setall

In this paper we will prove the set of highly non-trivial vanishing
conditions \eref{H0} to test the stability of our bundles. These conditions are a generalization of Hoppe's criterion \cite{hoppe}, used in \cite{Anderson:2007nc} to prove stability in the case of cyclic Calabi-Yau manifolds. In the cyclic case, the conditions in \eref{H0} are sufficient for stability, while for general CICYs, the vanishing of these cohomologies is necessary, but no longer sufficient. None-the-less, the generalized Hoppe condition still provides an important check of stability.
As mentioned earlier, Eq.~\eref{H0} is equivalent to the same condition but written in terms of the dual bundle, which turns out to be technically simpler. Hence, in this section we prove that
\begin{equation}\label{H0wedgepV*}
H^0(X, \wedge^p V^\star) = 0, \qquad p=1,\ldots,\rk(V) - 1\ .
\end{equation}
As mentioned previously, the full proof of stability of our bundles will appear in Ref.~ \cite{stabpaper}. 

We shall prove that condition \eref{H0wedgepV*} is satisfied in two steps. First, we demonstrate that the vanishing of certain ambient space cohomologies (given in \eref{wedgevdual=0}) associated to a Koszul resolution \eref{koszulA} guarantee that $H^0(X, \wedge^p V^\star)=0$.  After specifying these necessary cohomology groups, as a second step, we will show that they are in fact zero for all positive monads. This is accomplished by studying an exterior power sequence \eref{eagon} on $\cA$. We derive conditions \eref{cond1}, \eref{cond2}, and \eref{cond3} which hold for all the bundles in our classification.

\subsection{Step 1: Using the Koszul Sequence}
Let us begin with the Koszul resolution for $\wedge^p V$, which, from Eq.~\eref{koszulA}, reads
\begin{equation}
0 \to \wedge^K \cN^\star  \otimes \wedge^p \cV^\star  \to 
  \wedge^{K-1} \cN^\star  \otimes \wedge^p \cV^\star  \to \ldots
  \to \wedge^p \cV^\star  \to \wedge^p \cV^\star |_X \to 0 \ .
\end{equation}
In the above, $K$ is the co-dimension of the CICY $X$ and $\wedge^p \cV^\star $
is defined on the ambient product projective space $\cA$ while 
$\wedge^p \cV^\star |_X = \wedge^p V^\star $ lives on $X$. The normal bundle,
$\cN$ of $X$ in $\cA$ has been defined in Eq.~\eqref{normalbundle}.

\def\pvd{\wedge^p \cV^\star }
\def\pvdx{\wedge^p V^\star }
\newcommand{\wedgen}[1]{\wedge^{#1}\cN^\star }

We can break this long exact sequence into $K$ inter-related short exact 
ones by introducing $K-1$ (co-)kernels $Q_i$ such that
\begin{equation}
\begin{array}{lllllllll}
0 & \to &\wedgen{K}\otimes\pvd & \to & \wedgen{K-1}\otimes \pvd & 
  \to & Q_1  & \to & 0 \\
0 & \to &Q_1 & \to & \wedgen{K-2}\otimes \pvd & \to & Q_2  & \to & 0 \\
&&&&\vdots&&&&\\
0 & \to &Q_{K-2} & \to & \cN^\star  \otimes \pvd & \to & Q_{K-1}  & \to & 0 \\
0 & \to &Q_{K-1} & \to & \pvd & \to & \pvdx  & \to & 0 \\
\end{array}
\end{equation}
Now, each of the above short exact sequences induces a long exact sequence
in cohomology. Adopting the convention that $Q_0 := \wedgen{K}\otimes\pvd$ and
$Q_K := \pvdx$ (keeping in mind that $\wedgen{j} \simeq \cO_{\cA}$ for $j=0$),
the $j$-th long exact sequence takes the form
\begin{equation}
\begin{array}{lllllllll}
0 & \to & H^0(\cA,Q_{j-1}) & \to & H^0(\cA,\wedgen{K-j}\otimes \pvd)& 
    \to & H^0(\cA,Q_j)  & \to & \\
  & \to & H^1(\cA,Q_{j-1}) & \to & H^1(\cA,\wedgen{K-j}\otimes \pvd)& 
    \to & H^1(\cA,Q_j)  & \to & \\
  &&&&\vdots&&&&\\
  & \to & H^{K+2}(\cA,Q_{j-1}) & \to & H^{K+2}(\cA,\wedgen{K-j}\otimes \pvd)& 
    \to & H^{K+2}(\cA,Q_j)  & \to & \\
  & \to & H^{K+3}(\cA,Q_{j-1}) & \to & H^{K+3}(\cA,\wedgen{K-j}\otimes \pvd)& 
    \to & H^{K+3}(\cA,Q_j)  & \to & 0  \ . \\
\end{array}
\end{equation}
We have used the fact that $\cA$ is of dimension $K+3$ since $X$ has 
co-dimension $K$ and hence the highest cohomology group is $K+3$.
To ensure vanishing of $H^0(X, \pvdx)$ it suffices to have $H^0(\cA, \pvd)$
and $H^1(\cA, Q_{K-1})$ be zero. The latter vanishes, in turn, if
$H^1(\cA, \cN^\star  \otimes \pvd)$ and $H^2(\cA, Q_{K-2})$ are both zero. Thus arguing inductively, it is sufficient (though not necessary) for the vanishing of $H^0(X, \pvdx)$ that 
\begin{equation}\label{wedgevdual=0}
H^j(\cA, \wedgen{j}\otimes \pvd) = 0 \qquad \mbox{ for }
j = 0, \ldots, K \ .
\end{equation}
Indeed, these constitute $K+1$ vanishing conditions. For $j>K$,
$\wedgen{j} = 0$ since $\cN$ by definition is rank $K$ and the 
cohomologies are zero automatically. 

\subsection{Step 2: Using the Exterior Power Sequence}
How can we demonstrate that \eref{wedgevdual=0} is satisfied?
We recall the definition~\eqref{ambmonad} of the monad on the ambient space, whose dual is given by
\[
\sseq{\cC^\star }{\cB^\star }{\cV^\star } \ .
\]
The $p$-th exterior power of $\cV^\star $ can be extracted from the exterior-power sequence
\eref{eagon}, which here reads
\begin{equation}\label{ENwedgepd}
0 \to S^p \cC^\star  \to S^{p-1}\cC^\star  \otimes \cB^\star  \to \ldots \to
  \cC^\star  \otimes \wedge^{p-1} \cB^\star  \to  \wedge^{p} \cB^\star  \to \pvd \to 0 \ .
\end{equation}
By $S^j$ we denote the $j$-th symmetric tensor power. We can tensor this sequence by $\wedgen{j}$ for $j = 0, \ldots, K$. Each of the resulting $K+1$ sequences can be broken up into $p$ short exact ones, by introducing
(co-)kernels $q^j_i$, where $i=0,\ldots ,p-1$ and $j=0,\ldots ,K$. This leads to
\begin{equation}\label{q-kernels}
\begin{array}{lllllllll}
0 & \to &\wedgen{j} \otimes S^p \cC^\star  & \to & 
  \wedgen{j} \otimes S^{p-1}\cC^\star  \otimes \cB^\star  & 
  \to & q_1^j  & \to & 0 \\
0 & \to &q_1^j & \to & \wedgen{j} \otimes S^{p-2} \cC^\star  \otimes \wedge^2 \cB^\star  
  & \to & q_2^j  & \to & 0 \\
&&&&\vdots&&&&\\
0 & \to &q_{p-2}^j & \to & \wedgen{j} \otimes \cC^\star  \otimes \wedge^{p-1} \cB^\star  
  & \to & q_{p-1}^j & \to & 0 \\
0 & \to &q_{p-1}^j & \to & \wedgen{j} \otimes \wedge^{p} \cB^\star  & \to & 
  \fbox{$\wedgen{j} \otimes \pvd$} & \to & 0 \ , \\
\end{array} 
\end{equation}
where we have boxed the term whose $j$-th cohomology group on $\cA$ needs to
vanish.

Next, we consider the cohomology associated to \eref{q-kernels}. From Kodaira vanishing on $\cA$ (Eq.~\eref{kodaira-Amb}), a negative bundle $\cL^{\star}$ satisfies the vanishing conditions
\begin{equation}\label{kodaira-A}
H^{m}(\cA, \cL^\star) = 0 \mbox{ unless } m = \dim(\cA) = K+3\; .
\end{equation}
Our bundles $\cB^\star$, $\cC^\star$ as well as their tensors and powers are of course direct sums of strictly
negative bundles, and hence obey \eref{kodaira-A}. Each of the short exact sequences in Eq.~\eref{q-kernels} induces a long exact 
sequence in cohomology which are intertwined by the (co-)kernels. It will be helpful to consider \eref{q-kernels} and its cohomology for each value of $p$ individually. The results are immediate for the first two cases under consideration.

For $p=1$, we quickly see that $H^m(\cA, \wedgen{j} \otimes \pvd) = 0$
for $m = 0, \ldots, K+1$ as these are all sandwiched between two vanishing 
terms, namely $H^{m}(\cA, \wedgen{j} \otimes S^{p-1}\cC^\star  \otimes \cB^\star )$ and 
$H^{m+1}(\cA, \wedgen{j} \otimes S^p \cC^\star )$. Thus \eref{wedgevdual=0} is
automatically satisfied for $p=1$.
Similarly, for $p=2$, $H^m(\cA, \wedgen{j} \otimes \pvd) = 0$
for $m = 0, \ldots, K$, again satisfying Eq.~\eref{wedgevdual=0}. 

For longer exterior power sequences the result requires a little more analysis. For $p=3$, $H^m(\cA, \wedgen{j} \otimes \pvd)$ vanishes automatically only
for $m=0, \ldots, K-1$, one short of the upper bound of $j$ required in 
Eq.~\eref{wedgevdual=0}. Nevertheless, we find the equivalence 
$H^K(\cA, \wedgen{j} \otimes \pvd) \simeq H^{K+2}(\cA, q_1^j)$, and the latter cohomology group 
resides in the four-term exact sequence
\begin{equation}
0 \to H^{K+2}(\cA, q_1^j) \to 
H^{K+3}(\cA, \wedgen{j} \otimes S^{3}\cC^\star )
\stackrel{g}{\longrightarrow}
H^{K+3}(\cA, \wedgen{j} \otimes S^{2}\cC^\star  \otimes \cB^\star ) \to
H^{K+3}(\cA, q_1^j) \to 0 \ .
\end{equation}
The single case which remains to be checked is $j=K$.
It was argued in Appendix B of Ref.~\cite{Anderson:2007nc} that on the ambient
space, the map $g$ above, induced from the defining map of the monad
(which we recall, by construction, is generic), is also generic. Therefore,
if $g$ is injective, then the requisite term $H^{K+2}(\cA, q_1^j)$ vanishes. Injectivity simply requires that
\begin{equation}\label{cond1}
h^{K+3}(\cA, \wedgen{K} \otimes S^{3}\cC^\star )
\le
h^{K+3}(\cA, \wedgen{K} \otimes S^{2}\cC^\star  \otimes \cB^\star ) \ .
\end{equation}

At last, for the final case of $p=4$, $H^m(\cA, \wedgen{j} \otimes \pvd)$ vanishes automatically only
for $m=0, \ldots, K-2$, two short of the upper bound for $j$ in Eq.~\eref{wedgevdual=0}. However, we have that
$H^{K-1}(\cA, \wedgen{j} \otimes \pvd) \simeq H^{K+2}(\cA, q_1^j)$
the latter of which resides in a four-term exact sequence
\begin{equation}
0 \to H^{K+2}(\cA, q_1^j) \to 
H^{K+3}(\cA, \wedgen{j} \otimes S^{4}\cC^\star )
\stackrel{g_1}{\longrightarrow}
H^{K+3}(\cA, \wedgen{j} \otimes S^{3}\cC^\star  \otimes \cB^\star ) \to
H^{K+3}(\cA, q_1^j) \to 0 \ ;
\end{equation}
The relevant case $j=K-1$. As before, the map $g_1$ is generic and the requisite term $H^{K+2}(\cA, q_1^j)$
vanishes if $g_1$ is injective, or if
\begin{equation}\label{cond2}
h^{K+3}(\cA, \wedgen{K-1} \otimes S^{4}\cC^\star ) \le
h^{K+3}(\cA, \wedgen{K-1} \otimes S^{3}\cC^\star  \otimes \cB^\star ) \ .
\end{equation}
Similarly, we have that
$H^{K}(\cA, \wedgen{j} \otimes \pvd) \simeq H^{K+2}(\cA, q_2^j)$ and the latter resides in
\begin{equation}\label{hk3-1}
0 \to H^{K+2}(\cA, q_2^j) \to 
H^{K+3}(\cA, q_1^j)
\stackrel{g_2}{\longrightarrow}
H^{K+3}(\cA, \wedgen{j} \otimes S^{2}\cC^\star  \otimes \wedge^2 \cB^\star ) \to
H^{K+3}(\cA, q_2^j) \to 0 \ ,
\end{equation}
where we need to focus on the case $j=K$.
The cohomology $H^{K+3}(\cA, q_1^j)$ again resides in a four-term exact sequence
\begin{equation}\label{hk3-2}
0 \to H^{K+2}(\cA, q_1^j) \to 
H^{K+3}(\cA, \wedgen{j} \otimes S^{4}\cC^\star  )
\stackrel{h}{\longrightarrow}
H^{K+3}(\cA, \wedgen{j} \otimes S^{3}\cC^\star  \otimes \cB^\star ) \to
H^{K+3}(\cA, q_1^j) \to 0 \ .
\end{equation}
The maps $g_2$ and $h$ are generic, as before. Therefore, the cohomology at the end of Eq.~\eref{hk3-2}, 
$H^{K+3}(\cA, q_1^K) \simeq \coker(h)$ (which also appears as the second term of \eref{hk3-1})
has dimension $h^{K+3}(\cA, \wedgen{K} \otimes S^{3}\cC^\star  \otimes \cB^\star ) -
h^{K+3}(\cA, \wedgen{K} \otimes S^{4}\cC^\star  )$. For injectivity of $g_2$, this dimension should not exceed
$h^{K+3}(\cA, \wedgen{K} \otimes S^{2}\cC^\star  \otimes \wedge^2 \cB^\star )$, so we have the condition
\begin{equation}\label{cond3}
h^{K+3}(\cA, \wedgen{K} \otimes S^{3}\cC^\star  \otimes \cB^\star ) 
- h^{K+3}(\cA, \wedgen{K} \otimes S^{4}\cC^\star  )
\le
h^{K+3}(\cA, \wedgen{K} \otimes S^{2}\cC^\star  \otimes \wedge^2 \cB^\star ) \ .
\end{equation}
This condition then guarantees the vanishing of $H^{K+3}(\cA, q_1^K)$ and subsequently that of $H^{K}(\cA, \wedgen{K} \otimes \pvd)$.

We need not consider cases with $p>4$ since our bundles are maximally of rank 5.
In summary then, the conditions \eref{cond1}, \eref{cond2} and \eref{cond3} suffice to 
guarantee Eqs.~\eref{wedgevdual=0} and hence our main claim, Eq.~\eref{H0wedgepV*}. 
These conditions on the ambient space cohomology can be readily checked algorithmically using the Bott formula \eref{bott} and the K\"unneth formula \eref{kunneth1}. We have done so for all our
positive monads using computer scans and find these conditions are always satisfied.

In conclusion, for all positive monad bundles $V$,  $H^0(X, \wedge^p V^\star) = 0$ for $p=1,\ldots,\rk(V) - 1$. This concludes our non-trivial check of stability.


\section{Computing the Particle Spectrum}
\subsection{Bundle Cohomology}
While computing the full cohomology of monad bundles is generally a difficult task, it will become clear in the following that significant simplifications arise for positive monads. This computational advantage is of course one of the motivations to consider positive monads and it will lead to a number of general statements about their cohomology.

\subsubsection{Number of Families and Anti-families in $H^1(X,V)$ and $H^1(X,V^\star)$}
The defining short exact sequence~\eqref{monad} of the monad bundle $V$ induces the long exact sequence
\beq
\ba{lllllll}
 0&\to&H^0(X,V)&\to&H^0(X,B)&\to&H^0(X,C)\\
&\to&H^1(X,V)&\to&H^1(X,B)&\to&H^1(X,C)\\
&\to&H^2(X,V)&\to&H^2(X,B)&\to&H^2(X,C)\\
&\to&H^3(X,V)&\to&H^3(X,B)&\to&H^3(X,C)\to 0
\ea
\eeq 
Since both $B$ and $C$ are sums of positive line bundles we know from Kodaira vanishing that the cohomologies $H^q(X,C)=H^q(X,B)=0$ for all $q>0$. The above long exact sequence then immediately implies that $H^2(X,V)=0$. In the previous Section we have already shown that $H^0(X,V)=H^3(X,V)=0$ always, so that the only non-vanishing cohomology of positive monads is $H^1(X,V)$. The dimension $h^1(X,V)$ of this first cohomology can then be calculated from the index theorem~\eqref{indtheorem} or indeed the above long exact sequence. In summary, one finds
\begin{equation}
 h^1(X,V)=h^0(X,C)-h^0(X,B)=-{\rm ind}(V)\; ,\quad h^q(X,V)=0\mbox{ for } q\neq 1\; .
\end{equation} 
This means that the number of anti-families always vanishes and that the number of families can easily be read off from the index in Figs.~\ref{f:c3pos} and \ref{f:c3pos345}.
The absence of vector-like pairs of families might be considered an attractive feature and is certainly a pre-requisite for compactifications with the exact standard model spectrum. We stress that this property is directly linked to the property of positivity and will not generally hold if we allowed zero or negative integer entries in the line bundles defining the monad.

\def\cv2s{\wedge^2 {\cal V}^\star }

\subsubsection{Computing $H^1(X, \wedge^2 V^\star)$ and Number of Higgs Multiplets}
For ${\rm SU}(3)$ bundles we have $V\simeq \Lambda^2 V^\star$ and, hence,  the cohomology groups $H^1(X,\wedge^2 V)$ and $H^1(X, \wedge^2 V^\star)$ contain no new information. However, for ${\rm SU}(4)$ and ${\rm SU}(5)$ this is not the case and we have to perform another calculation. In the case of rank four, $\wedge^2V\simeq \wedge^2V^\star$, so that $H^1(X,\wedge^2 V)\simeq H^1(X, \wedge^2 V^\star)$. For rank five the situation is less trivial, but from Eq.~\eqref{indtheorem2} we know that $h^1(X,\wedge^2 V)$ and $h^1(X,\wedge^2V^\star)$ are related by the index, ${\rm ind}(V)$, of $V$. Hence, in both the rank four and five cases it is enough to compute one of $H^1(X,\wedge^2 V)$ and $H^1(X,\wedge^2V^\star)$ and, in the following, we will opt for $H^1(X,\wedge^2V^\star)$.

To calculate this cohomology, we proceed as in Section \ref{s:stable}. Since the arguments therein
were stated for general anti-symmetric power $p$, it is instructive to be more explicit here.
We start by writing down the Koszul resolution~\eref{koszulA} for $\wedge^2V^\star$ which is given by
\beq\label{koszul-cv2s}
0 \to \cv2s \otimes \wedge^K \cN^\star  \to \cv2s \otimes \wedge^{K-1} \cN^\star 
\to \ldots \to \cv2s \otimes \cN^\star  \to \cv2s \to \wedge^2V^\star \to 0 \ .
\eeq
Recall that $K$ is the co-dimension of the CICY $X$ embedded in the ambient space $\cA$ and $\cN$ is the normal bundle~\eqref{normalbundle} of $X$ in $\cA$. As a first step we will now derive vanishing theorems for the cohomologies of the bundles $\wedge^2V^\star \otimes \wedge^j\cN^\star$ which appear in the above Koszul sequence. 
To do this, we start the exact sequence for antisymmetric products of bundles from
\eref{eagon}:
\beq\label{wedge2V*}
0 \to S^2 \cC^\star  \to \cC^\star  \otimes \cB^\star  \to \wedge^2 \cB^\star  \to \cv2s \to 0 \ ,
\eeq
which is induced from the dual sequence 
\beq
\sseq{\cC^\star }{\cB^\star }{\cV^\star }\; .
\eeq
We can then tensor \eref{wedge2V*} by $\wedge^j \cN^\star $ for $j=0,\ldots, K$ and break the resulting 4-term exact sequence into two short exact sequences
\beq\ba{l}
\sseq{S^2 \cC^\star  \otimes \wedge^j \cN^\star }
     {\cC^\star  \otimes \cB^\star  \otimes \wedge^j \cN^\star }{Q_j} \ ;\\
\sseq{Q_j}{\wedge^2 \cB^\star  \otimes \wedge^j \cN^\star }
     {\wedge^2 \cV^\star  \otimes \wedge^j \cN^\star } \ ;
\ea \qquad
j = 0, \ldots, K \ ,
\eeq
where $Q_j$ are approriate (co)kernels. This induces two inter-related
long exact sequences in cohomology on $\cA$ which are given by

\def\cjn{{\cA, S^2 \cC^\star  \otimes \wedge^j \cN^\star }}
\def\cbjn{{\cA, \cC^\star  \otimes \cB^\star \otimes \wedge^j \cN^\star }}
\def\bjn{{\cA, S^2 \cB^\star  \otimes \wedge^j \cN^\star }}

\beq\label{Hwedege2V*}
\ba{llllllllll}
0
 &\to& \cancelto{0}{H^0(\cjn)} &\to& \cancelto{0}{H^0(\cbjn)} &\to& H^0(\cA, Q_j) &\to&\\
 &\to& \cancelto{0}{H^1(\cjn)} &\to& \cancelto{0}{H^1(\cbjn)} &\to& H^1(\cA, Q_j) &\to&\\
 &\to& &&\vdots &&  &\to& \\
 &\to& \cancelto{0}{H^{K+2}(\cjn)} &\to& \cancelto{0}{H^{K+2}(\cbjn)} &\to& H^{K+2}(\cA, Q_j) &\to&\\
 &&&&&&&&\\ 
 &\to& H^{K+3}(\cjn) &\to& H^{K+3}(\cbjn) &\to& H^{K+3}(\cA, Q_j) &\to& 0 \ ;\\ \\
0
 &\to& H^0(\cA, Q_j) &\to& \cancelto{0}{H^0(\bjn)} &\to&  {H^0(\cA, \cv2s\otimes \wedge^j \cN^\star )} &\to&\\
 &\to& H^1(\cA, Q_j) &\to& \cancelto{0}{H^1(\bjn)} &\to&  {H^1(\cA, \cv2s\otimes \wedge^j \cN^\star )} &\to&\\
 &\to& &&\vdots &&  &\to& \\
 &\to& H^{K+2}(\cA, Q_j) &\to& \cancelto{0}{H^{K+2}(\bjn)} &\to& 
 		{H^{K+2}(\cA,\cv2s\otimes \wedge^j \cN^\star  )} &\to& \\
 &&&&&&&&\\ 
 &\to& H^{K+3}(\cA, Q_j) &\to& H^{K+3}(\bjn) &\to& H^{K+3}(\cA,\cv2s\otimes \wedge^j \cN^\star  ) &\to&
0 \ .
\ea
\eeq
Note that since $X$ is of codimension $K$, the ambient space has
dimension $K+3$ and hence there are no non-vanishing cohomology groups
above $H^{K+3}$. Moreover,
the bundles $\cN^\star$, $\cB^\star$ and $\cC^\star$ as well as their various tensor and wedge products are all negative and, 
hence, all their cohomologies except the highest one, namely $K+3$, vanish by \eref{kodaira-A};
we have marked this explicitly
in Eq.~\eref{Hwedege2V*}.

Therefore, the sequences~\eref{Hwedege2V*} immediately imply that for all $j$,
\beq\label{wedege2V*vanish}
\ba{lll}
H^i(\cA, Q_j) = 0 \ , && i=0, \ldots, K+1\ ; \\
H^i(\cA, \cv2s\otimes \wedge^j \cN^\star ) \simeq H^{i+1}(\cA, Q_j)  = 0 \ , && i=0, \ldots, K \ ; \\
H^{K+1}(\cA, \cv2s\otimes \wedge^j \cN^\star ) \simeq H^{K+2}(\cA, Q_j)  &&
\ea\eeq
as well as two 4-term exact sequences:
\beq\label{posHwedege2V*}
\ba{l}
0 \to H^{K+2}(\cA, Q_j) \to H^{K+3}(\cjn) \stackrel{g}{\longrightarrow} H^{K+3}(\cbjn) \to H^{K+3}(\cA, Q_j) \to 0
\ ;
\\ \\
0 \to H^{K+2}(\cA, \cv2s\otimes \wedge^j \cN^\star ) \to H^{K+3}(\cA, Q_j) \to H^{K+3}(\bjn) \to H^{K+3}(\cA, \cv2s\otimes \wedge^j \cN^\star ) \to
0 \ .
\ea
\eeq
In Eq.~\eref{posHwedege2V*} we have introduced a map $g$ which is induced from the defining map $f$ of the monad in Eq.~\eref{defV}.  
As in the previous subsection, $g$ is generic and thus has maximal rank. The top
sequence then implies that $H^{K+2}(\cA, Q_j)=0$ and, hence, by Eq.~\eref{wedege2V*vanish},
$H^{K+1}(\cA, \cv2s\otimes \wedge^j \cN^\star )$ vanishes as well.
To summarise then,  we find the vanishing cohomology groups
\beq\label{Hiwedgej}
H^{i}(\cA, \cv2s\otimes\wedge^j \cN^\star )=0 \ , \qquad
\forall~i = 0, \ldots K+1, \ j = 0, \ldots, K \ .
\eeq

Equipped with these results, we can re-examine the Koszul sequence~\eref{koszul-cv2s}. It has $K+2$ terms
and we can break it up into $K$ short exact sequences, introducing (co)kernels much like we did above. Then, the vanishing of the cohomology groups
\beq\label{Hj+1wedgej}
H^{j+1}(\cA, \wedge^2\cV^\star  \otimes \wedge^j \cN^\star ) = 0 \ , \qquad
\forall~j=0, \ldots, K \ ,
\eeq
which represent a subset of the vanishing theorems~\eqref{Hiwedgej}, implies that
\beq\label{H1wedge=0}
H^1(X, \wedge^2 V^\star ) = 0 \ .
\eeq

We emphasize that the assumption of a generic map $f$, which defines the monad in \eref{monad}, 
is crucial to arrive at this result. For rank four bundles with low-energy gauge group ${\rm SO}(10)$ it implies (see Table~\ref{t:spec}) that
\begin{equation}
 n_{10}=h^1(X,\wedge^2 V)=0\; ,
\end{equation}
and, hence, a vanishing number of Higgs multiplets. For rank five bundles with low-energy gauge group ${\rm SU}(5)$ we have  
\begin{equation}
 n_5=h^1(X,\wedge^2V^\star)=0\; ,\quad n_{\bar 5}=-{\rm ind}(V)\; ,
\end{equation}
where Eq.~\eqref{indtheorem2} has been used. This means the number of ${\bf 10}$ and $\bar{\bf 5}$ representations  is the same, forming complete ${\rm SU}(5)$ families and there are no vector-like pairs of ${\bf 5}$ and $\bar{\bf 5}$ representations. The absence of Higgs multiplets in the ${\rm SO}(10)$ and ${\rm SU}(5)$ models is a phenomenologically problematic feature which was already observed in Ref.~\cite{Anderson:2007nc}. There, it has also been shown that the number of Higgs multiplets can be non-zero once the assumption of a generic map $f$ is dropped. A similar situation was encountered in \cite{Bouchard:2005ag}.

We expect a similar bundle-moduli dependence of the spectrum, as first discussed in \cite{Donagi:2004qk},
for the more general class of models considered in this paper. 
It remains a matter of a more detailed analysis, focusing on physically promising models within our classification, to decide if a realistic particle spectrum can be obtained from such a mechanism.

\subsubsection{Singlets and $H^1(X, V \otimes V^\star )$}
Finally, we need to calculate the number of gauge group singlets which correspond to the cohomology
$H^1(X, \text{ad}(V)) = H^1(X, V \otimes V^\star )$. We begin by tensoring the defining sequence \eqref{dualV} for $V^\star $ by $V$. This leads to a new short exact sequence
\beq
0 \to C^\star  \otimes V \to B^\star  \otimes V \to V^\star  \otimes V \to 0 \ .
\eeq
One can produce two more short exact sequences by multiplying
\eqref{dualV} with $B$ and $C$. Likewise, three short exact
sequences can be obtained by multiplying the original sequence~\eqref{monad}
for $V$ with $V^\star $, $B^\star $ and $C^\star $. The resulting six
sequences can then be arranged into the following web of three horizontal
sequences $h_{I}$, $h_{II}$, $h_{III}$ and three vertical ones
$v_I$, $v_{II}$, $v_{III}$.
\beq\ba{cccccccccl}
&&0&&0&&0&&& \\
&&\downarrow&&\downarrow&&\downarrow&&& \\
0&\to& C^\star  \otimes V &\to& B^\star  \otimes V &\to& V^\star  \otimes V &\to&0
\qquad &h_{I} \\
&&\downarrow&&\downarrow&&\downarrow&&& \\
0&\to& C^\star  \otimes B &\to& B^\star  \otimes B &\to& V^\star  \otimes B &\to&0
\qquad &h_{II} \\
&&\downarrow&&\downarrow&&\downarrow&&& \\
0&\to& C^\star  \otimes C &\to& B^\star  \otimes C &\to& V^\star  \otimes C &\to&0
\qquad &h_{III}  \\
&&\downarrow&&\downarrow&&\downarrow&&& \\
&&0&&0&&0&&& \\
&&v_I&&v_{II}&&v_{III}&&& \\
\ea
\eeq
The long exact sequence in cohomology induced by $h_{I}$ reads
\bea\label{VVseq}
0 &\to& H^0(X,C^\star  \otimes V) \to H^0(X,B^\star  \otimes V)
\to H^0(X, V^\star  \otimes V)\nn\\
&\to&H^1(X,C^\star  \otimes V) \to
 H^1(X, B^\star  \otimes V)\to \fbox{\mbox{$H^1(X,V^\star  \otimes V)$}}\nn \\
&\to& H^2(X,C^\star  \otimes V) \to \ldots\label{h1}
\eea
and we have boxed the term which we would like to compute. We will also need the
long exact sequences which follow from $v_{I}$ and $v_{II}$. They
are given by
\bea
 0&\to&H^0(X,C^\star \otimes V)\to H^0(X,C^\star \otimes B)\to H^0(X,C^\star \otimes C)\nn\\ 
  &\to&H^1(X,C^\star \otimes V)\to H^1(X,C^\star \otimes B)\to H^1(X,C^\star \otimes C)\nn\\
 &\to&H^2(X,C^\star \otimes V)\to H^2(X,C^\star \otimes B)\to H^2(X,C^\star \otimes C)\nn\\
&\to&H^3(X,C^\star \otimes V)\to H^3(X,C^\star \otimes B)\to H^3(X,C^\star \otimes C) \to 0 \ ;
 \label{v1}
 \\[0.3cm]
 0&\to&H^0(X,B^\star \otimes V)\to H^0(X,B^\star \otimes B)\to H^0(X,B^\star \otimes C)\nn\\ 
  &\to&H^1(X,B^\star \otimes V)\to H^1(X,B^\star \otimes B)\to H^1(X,B^\star \otimes C)\nn\\
 &\to&H^2(X,B^\star \otimes V)\to H^2(X,B^\star \otimes B)\to H^2(X,B^\star \otimes C)\nn \\
 &\to&H^3(X,B^\star \otimes V)\to H^3(X,B^\star \otimes B)\to H^3(X,B^\star \otimes C)\to 0 \ .
 \label{v2}
 \eea
To make progress we need information about the cohomologies of $B^\star\otimes B$, $C^\star\otimes C$ and $C^\star\otimes B$. For the cyclic cases (the CICYs with $h^{1,1}(X)=1$) discussed in Ref.~\cite{Anderson:2007nc} all line bundles $L$  on $X$ have vanishing middle cohomologies, that is $H^1(X,L)=H^2(X,L)=0$ and, hence, the same is true for $B^\star\otimes B$, $C^\star\otimes C$ and $C^\star\otimes B$. For the general case discussed here this is no longer necessarily true since $B^\star\otimes B$, $C^\star\otimes C$ and $C^\star\otimes B$ may contain ``mixed" line bundles with different sign or zero entries which may have non-vanishing middle cohomologies. This means in some cases there will not be sufficiently many zero entries in the above long exact sequences to compute $h^1(X,V\otimes V^\star)$ without additional input, for example about the rank of maps. 

However, a general formula can be derived for all monads satisfying
\begin{equation}
 H^1(X,C^\star\otimes C)=H^2(X,C^\star\otimes B)=0\; . \label{vancond}
\end{equation} 
Since we can compute all line bundle cohomologies we can explicitly check for each given example whether these conditions are actually satisfied. Let us focus on models where this is the case. Then the sequence~\eqref{v1} implies that $H^2(X,C^\star\otimes V)=0$ which means that \eqref{h1} breaks after the second line and this 6-term exact sequence implies:
 \beq
 h^1(X,V^\star \otimes V)=h^1(X,B^\star \otimes V)-h^1(X,C^\star \otimes V)
 +h^0(X,V^\star \otimes V) -h^0(X,B^\star \otimes V) + h^0(X, C^\star \otimes V) \; .
 \label{nsing0}
\eeq
In the above, we have used the fact that for any long exact sequence, whatever the number of terms, the total alternating sum of the dimensions of the terms vanishes.

We can apply a similar trick to the other 2 long exact sequences.
Using our assumptions $H^1(X,B^\star\otimes C)\simeq H^2(X,C^\star\otimes B)=0$ in the sequence~\eqref{v2} and 
$H^1(X,C^\star\otimes C)=0$ in the sequence~\eqref{v1} gives the two relations
\bea
\nn 
 h^1(X,B^\star \otimes V)- h^0(X,B^\star \otimes V)&=&h^0(X,B^\star \otimes C) -h^0(X,B^\star \otimes B)+h^1(X,B^\star\otimes B)\\
 \nn
 h^0(X,C^\star\otimes V)-h^1(C^\star\otimes V)&=&h^0(X,C^\star\otimes B)-h^0(X,C^\star\otimes C)-h^1(X,C^\star\otimes B)\; .
\eea
Inserting these into Eq.~\eqref{nsing0} and using the fact that for a stable $SU(n)$ bundle $V$,
$h^0(X,V \otimes V^\star ) = 1$ (see Section 4.2 of \cite{Anderson:2007nc}) gives the final result
\bea
 n_1=h^1(X,V^\star\otimes V)&=&h^0(X,B^\star\otimes C)-h^0(X,B^\star\otimes B)-h^0(X,C^\star\otimes C)\nonumber\\
 &&+h^0(X,C^\star\otimes B)-h^1(X,C^\star\otimes B)+h^1(X,B^\star\otimes B)+1 \label{n1res}
\eea
for the number of singlets. We emphasize that this is result is valid provided the monad satisfies the two conditions~\eqref{vancond}. 
In this case, Eq.~\eqref{n1res} allows an explicit calculation of the number of singlets from the known line bundle cohomologies.

As an example, we consider the manifold 
$\left[
\begin{array}
[c]{c}
1\\
3
\end{array}
\left|
\begin{array}
[c]{ccc}
2 \\
4
\end{array}
\right.  \right]  $, and the rank 4 monad bundle defined by
\beq
B=\cO_X(1,1)^{\oplus 6} \oplus \cO_X(2,1)^{\oplus 2}\; ,\quad C=\cO_X(2,3)^{\oplus2} \oplus \cO_X(3,1)^{\oplus2} \; .
\eeq
It can be checked from the known line bundle cohomologies that this bundle indeed satisfies the conditions~\eqref{vancond}. The number of singlets, calculated from Eq.~\eqref{n1res}, is then given by $n_1=241$.

For bundles which do not satisfy \eqref{vancond} other methods can be employed. In favourable cases, the cohomologies of $B^\star\otimes B$, $C^\star\otimes C$ and $C^\star\otimes B$ may have a different pattern of zeros which still allows the derivation of a formula for $n_1$ analogous to Eq.~\eqref{n1res} by combining appropriate parts of the sequences~\eqref{h1}, \eqref{v1} and \eqref{v2}. If this is not possible one has to resort to ambient space methods and Koszul resolutions in combination with our results for the ranks of maps in Leray spectral sequences. Here, we will not present such a calculation which is likely to be complicated and, if required at all, should probably be only carried out for physically promising models. However, we stress that all the necessary technology is available so that the number of singlets can, not just in principle but in practice, be obtained for all positive monads on favourable CICYs.


\section{Conclusions and Prospects}\setall
In this paper, we have analysed positive monad bundles with structure group ${\rm SU}(n)$ (where $n=3,4,5$) on favourable CICY manifolds in the context of $N=1$ supersymmetric compactifications of the $E_8\times E_8$ heterotic string. We have shown that the class of these bundles, subject to the heterotic anomaly condition, is finite and consists of $7118$ examples. More specifically, we find that these $7000$ or so monads are concentrated on only $36$ CICYs. All other of the $4500$ or so CICYs do not allow positive monads which satisfy the anomaly condition. 

As a highly non-trivial test for the stability of these bundles we have shown that $H^0(X,\wedge^p V^*)=0$ for $p=1, \ldots, \rk(V)-1$ for all our examples. A systematic stability proof will be presented in Ref.~\cite{stabpaper}. We have also shown how to calculate the complete particle spectrum for these models. In particular, we found that the number of anti-families always vanishes so that there are no vector-like family anti-family pairs present in any of the models. For low-energy groups ${\rm SO}(10)$ and ${\rm SU}(5)$ ($n=4,5$) the number of Higgs fields vanishes at generic points in the bundle moduli space. However, as was shown in Ref.~\cite{Anderson:2007nc}, for non-generic values of the bundle moduli, Higgs multiplets can arise. The details of this moduli-dependence of the spectrum (see Ref.~\cite{Donagi:2004qk}) 
have to be analysed for specific models, preferably focusing on physically promising examples. 
Furthermore, we have shown that the number of gauge singlets can be calculated, in many cases in terms of a generic formula, or else by applying more elaborate methods. 

Based on the results for the particle spectrum, we have scanned the $7118$ bundles imposing two rudimentary physical conditions. First, the number of families should equal $3k$ for some non-zero integer $k$, so there is a chance to obtain three families after dividing by a discrete symmetry of order $k$. In addition, the Euler number of the Calabi-Yau space should be divisible by $k$. It turns out that only $559$ out of the $7118$ bundles pass this basic test. If, in addition, one demands that the order $k$ of the symmetry does not exceed $13$ one is left with only $21$ models. 

This drastic reduction of the number of viable models due to a few basic physical constraints is not uncharacteristic and has been observed in the context of other string constructions~\cite{Gmeiner:2005vz,Gmeiner:2007zz}.
In our case, the main reason for this reduction is the relatively large values for the Euler characteristic of our models (roughly, a Gaussian distribution with a maximum at about $60$, see Fig.~\ref{f:c3pos}) in conjunction with the empirical fact that large discrete symmetries of Calabi-Yau manifolds are hard to find. In order to make this statement more precise a systematic analysis of discrete symmetries $\Gamma$ on CICYs $X$ (which lead to a smooth quotient $X/\Gamma$) has to be carried out and the results of this analysis have to be combined with the results of the present paper. We are planning to perform this explicitly in the near future. However, even in the absence of such a classification of discrete symmetries we find it likely that the vast majority of positive monads will fail to produce three-family models on $X/\Gamma$ given the large number of families on the ``upstairs" manifold $X$. 

These large numbers are, of course, directly related to the property of positivity. An obvious course of action is, therefore, to relax this condition and also allow zero or even slightly negative integers $b_i^r$ and $c_j^r$ in the definition~\eqref{monad} of the monad. The number of these non-positive monads is vastly larger than the number of positive ones and it turns out the distribution of their Euler characteristics is peaked at smaller values, as expected. Crucially, as will be shown in Ref.~\cite{stabpaper}, some of these non-positive monads are still stable and, hence, lead to supersymmetric models. We, therefore, believe that the generalisation to non-positive monads is a crucial step towards realistic models within this framework and work in this direction is underway~\cite{zeropaper}.

\section*{Acknowledgments}
The authors would like to expression our sincere gratitude to
Philip Candelas, Tristan H\"ubsch, Adrian Langer, 
Balazs Szendroi and Andreas Wisskirchen for many helpful discussions.
L.~A.~thanks the US NSF and the Rhodes Foundation for support. 
Y.-H.~H~is indebted to the UK STFC for an Advanced Fellowship as well as 
the FitzJames Fellowship of Merton College, Oxford. 
A.~L.~is supported by the EC 6th Framework Programme MRTN-CT-2004-503369.  

\appendix

\section{Notation and conventions}
Throughout the paper we will adhere to the following notations:\\[0.3cm]
\begin{tabular}{ll}
$X$ & Calabi-Yau threefold embedded in ambient space $\cA=\mathbb{P}^{n_1}\otimes\dots\otimes\mathbb{P}^{n_m}$ \\
$[q_j^r]_{j=1,\ldots ,K}^{r=1,\ldots ,m}$ & configuration matrix for co-dimension $K$ CICY in product of $m$ projective spaces\\
$\cO_\cA ({\bf k})$&product of line bundles $\cO_\cA (k^1)\otimes\dots\otimes\cO_\cA (k^m)$ on $\cA$\\ 
$\cO_X({\bf k})$&restriction of $\cO_\cA ({\bf k})$ to $X$\\
$\cN$ & Normal bundle of $X$ in $\cA$ \\
$TX$ & Tangent bundle of $X$; similarly, $T\cA$ is the tangent bundle
  of $\cA$\\ 
$V$ & Vector bundle on $X$, the dual bundle is denoted $V^\star $ \\
$B,C$ & Sum of line bundles $\bigoplus_i\cO_X({\bf b}_i)$ and $\bigoplus_a\cO_X({\bf c}_a)$ on $X$ \\
$\cV$ & Vector bundle on ambient $\cA$ which restricts to $V$ on $X$ \\
$\cB, \cC$ & Sums of line bundles $\bigoplus_i\cO_\cA({\bf b}_i)$ and $\bigoplus_a\cO_\cA({\bf c}_a)$ on $\cA$ \\
\end{tabular}
\section{Some Mathematical Preliminaries}
In this appendix, we collect some useful mathematical facts which will be of importance
throughout the paper. These can be found in standard references such as \cite{AG1,AG2,FH,hubsch}.
\paragraph{Serre Duality: }
For a vector bundle $V$ on a manifold $M$ of complex dimension $n$, Serre duality relates the
cohomology groups of $V$ with those of its dual as:
\beq\label{serre}
H^i(M,V) \simeq H^{n-i}(X, V^\star  \otimes K_M) \qquad i=0,1, \ldots, n \ ,
\eeq
where $K_M=\bigwedge^{n} TM^\star$ is the canonical bundle of $M$.
For a Calabi-Yau threefold $X$,  the canonical bundle $K_X$ is the trivial bundle $\cO_X$ and, hence, Serre duality takes the particularly simple form
\beq\label{serre-CY}
H^i(X,V) \simeq H^{3-i}(X, V^\star ) \qquad i=0,1,2,3 \ .
\eeq

\paragraph{Atiyah-Singer Index Theorem: }
For a unitary bundle $V$ on a Calabi-Yau threefold $X$, the index theorem relates the index, or the alternating sum of dimensions of the cohomology groups of $V$ with the characteristic classes of the bundle and the manifold:
\beq\label{AS}
\ind(V) = \sum\limits_{i=0}^3 (-1)^i h^i(X,V) = 
\int_X \ch(V) \wedge \td(X) = \frac12 \int_X  c_3(V)
\ ,
\eeq
where $\td(X)$ is the Todd class for the tangent bundle of $X$. Only in
the last equality have we used the fact the both $c_1(TX)$ and
$c_1(V)$ vanish.

\paragraph{Higher Exterior Powers: }
For $SU(n)$ bundles we have the equivalences
\beq\label{SUn}
\wedge^p V \simeq \wedge^q V^\star  \qquad  p+q=n \ 
\eeq
and the relation (see Appendix B of Ref.~\cite{Donagi:2004ia}),
\beq
c_3(\wedge^2 V) = (n-4) c_3(V) \ .
\eeq

\paragraph{The Bott Formula: }
The cohomology of line-bundles over a projective space $\mathbb{P}^n$ is given by a simple formula,
the so-called Bott formula (see, for example, Ref.~\cite{Distler:1987ee}), which dictates that
\beq\label{bott}
h^q(\IP^n, (\wedge^p T\IP^n) \otimes \cO_{\mathbb{P}^n}(k)) =
\left\{\ba{lll}
{k+n+p+1 \choose p}{k+n \choose n-p} & q = 0 & k>-p-1,\\
1 & q=n-p & k=-n-1,\\
{-k-p-1 \choose -k-n-1}{-k-n-2 \choose p} & q = n & k<-n-p-1,\\
0 & \mbox{otherwise} & \ .
\ea\right.
\eeq

\paragraph{K\"unneth formula: }
The K\"unneth formula gives the cohomology of bundles over direct product of
spaces. For a product $\cA=\mathbb{P}^{n_1}\otimes\dots\otimes\mathbb{P}^{n_m}$ of projective spaces and ${\bf k}=(k^1,\ldots ,k^m)$, it states that
\beq\label{kunneth1}
H^n(\cA, \cO_\cA({\bf k})) =
\bigoplus_{q_1+\ldots+q_m = n} H^{q_1}(\IP^{n_1},\cO_{\mathbb{P}^{n_1}}(k^1)) \times
\ldots \times H^{q_m}(\IP^{n_m},\cO_{\mathbb{P}^{n_m}}(k^m)) \ ,
\eeq

\paragraph{Kodaira Vanishing Theorem: }
For positive line bundle $L$ on a Kahler manifold $M$ the Kodaira vanishing theorem states that 
\beq\label{kodaira}
H^{q}(M,L \otimes K_{M}) = 0 \quad \forall~q>0 \ ,
\eeq
where $K_{M}$ is the canonical bundle on $M$. For a Calabi-Yau manifold, $X$,
$K_{X}$ is trivial and therefore the only non-vanishing cohomology for
a positive line bundle, $L$, on $X$ is $H^{0}(X,L)$.
On the ambient space $\cA$ it is useful to look at the the Serre dual of Eq.~\eref{kodaira}. For positive line bundles $\cL$ on $\cA$ Eq.~\eref{serre} this leads to $H^q(\cA, \cL \otimes K_{\cA}) \simeq H^{\dim(\cA)-q}(\cA,\cL^\star  \otimes K_{\cA}^\star \otimes K_{\cA})$. The canonical bundle $K_{\cA}$ and its dual tensor to $\cO_{\cA}$ and we have the important fact that
\begin{equation}\label{kodaira-Amb}
H^{q}(\cA, \cL^\star) = 0 \mbox{ unless } q = \dim(\cA).
\end{equation}

\paragraph{Koszul Resolution}
The standard method of computing the cohomology of a vector bundle
$V= \cV|_X$  obtained by restricting the bundle (or sheaf) $\cV$  on the ambient space $\cA$ to
the variety $X$ is the so-called {\it Koszul Resolution} of
$V$. In general, if $X$ is a smooth hypersurface of co-dimension $K$, which is the zero locus of a holomorphic section $s$ of the bundle $\cN$, then the following exact sequence exists:
\beq\label{koszulA}
0 \to \cV \otimes \wedge^K \cN^\star  \to \cV \otimes \wedge^{K-1} \cN^\star 
\to \ldots \to \cV \otimes \cN^\star  \to \cV \to V \to 0 \ .
\eeq
Thus, if the cohomology of the bundles $\wedge^j \cN^\star  \otimes \cV$ are known on the ambient space, we can use the Koszul sequence to determine the cohomology of $V$. We recall that for a CICY, the normal bundle is given in terms of the configuration matrix, as in Eq.~\eqref{normalbundle}.

\paragraph{Exterior-Power Sequence: }
Given a short exact sequence of vector bundles $A$, $B$ and $C$ on any manifold:
\[
\sseq{A}{B}{C} \ ,
\]
there exists a long exact sequence for the $p$-th exterior power of $C$, derivable from a so-called
Eagon-Northcott complex. This sequence reads:
\begin{equation}\label{eagon}
0 \to S^p A \to S^{p-1}A \otimes B \to \ldots \to
 A \otimes \wedge^{p-1} B \to  \wedge^{p} B \to \wedge^p C \to 0 \ .
\end{equation}

\section{More on CICYs}\label{a:cicy}
We have introduced basic facts about CICYs in the main text. In this appendix, we present some more detailed properties relevant to our investigation. Many of these are standard results which can be found, for example, in Ref.~ \cite{hubsch} but we also discuss some new aspects, in particular the redundancy in the CICY list. 
\subsection{Chern Classes and Intersection Form}
We focus on a class of CICYs $X$, defined as the common zero locus of $K$ polynomials in an ambient space $\cA=\mathbb{P}^{n_1}\otimes\dots\otimes\mathbb{P}^{n_m}$ with $m$ projective factors of dimension $n_r$.. This CICY is characterised by a configuration matrix $[q_j^r]_{j=1,\ldots ,K}^{r=,\ldots ,m}$, as in Eq.~\eqref{cy-config}, where $q_j^r$ denotes the degree of the $j^{\rm th}$ polynomial in the variables of the $r^{\rm th}$ projective space. These degrees are subject to the complete intersection condition~\eqref{ci} and the condition~\eqref{cy-deg} which ensures the vanishing of the first Chern class $c_1(TX)$. Integration over $X$ can be reduced to integration over the ambient space $\cA$ using the formula
\beq\label{integration}
\int_X \cdot = \int_{\cA} \mu \wedge \cdot \ , \qquad
\mu := \wedge_{j=1}^K \left( \sum_{r=1}^m q^j_r J_r \right) \ .
\eeq
In this way, one can compute the triple intersection numbers 
\begin{equation}
 d_{rst}=\int_XJ_r\wedge J_s\wedge J_t \label{drst}
\end{equation}
where $J_r$ are the Kahler forms of the ambient space projective factors $\mathbb{P}^{n_r}$.
The Chern classes are given as simple functions of the entries in the configuration matrix \cite{hubsch}. The total Chern class can be expanded in terms of the ambient space Kahler forms as
\beq
c(TX) = c_1^r(TX) J_r + c_2^{rs}(TX) J_r J_s + c_3^{rst}(TX) J_r J_s J_t \ ,
\eeq
where
\bea\label{chernX}
c_1^r(TX) &=& 0\\
c_2^{rs}(TX) &=& \frac12 \left[ -\delta^{rs}(n_r + 1) + 
  \sum_{j=1}^K q^r_j q^s_j \right] \\
c_3^{rst}(TX) &=& \frac13 \left[\delta^{rst}(n_r + 1) - 
  \sum_{j=1}^K q^r_j q^s_j q^t_j \right] \ .
\eea
The second Chern class should be expressed as $c_2(TX)=c_{2r}(TX)\nu^r$ relative to a basis $\nu^r$ of $H^4(X,\mathbb{Z})$, as defined in Eq.~\eqref{nur}. The conversion from the coefficients $c_2^{rs}(TX)$ above can be accomplished by contraction with the intersection numbers
\begin{equation}
 c_{2r}(TX)=d_{rst}c_2^{st}(TX)\; .
\end{equation} 
Similarly, the Euler number $\chi (X)$ is obtained from
\begin{equation}
\chi (X)=d_{rst}c_3^{rst}(TX)\; .
\end{equation}

%
\subsection{Hodge Numbers}
We wish to know the full topological data of $X$ including the
Hodge numbers $h^{1,1}(X)$ and $h^{2,1}(X)$, whose difference, by the Index
Theorem \eref{AS}, is the Euler number $\chi(X)$:
\beq
h^{1,1}(X) - h^{2,1}(X) = \frac12\, \chi(X) \ .
\eeq
Therefore, it suffices to compute either one of these two Hodge numbers. This calculation is the subject of Ref.~\cite{Green:1987cr} and it this turns out to be much more involved than calculating the Euler number.
While this paper explains the basic method, sadly, the actual data for these Hodge numbers seems to have been lost. 
Both for the purpose of reconstructing this data and because related techniques can be applied to monad bundles it is useful to review the methods of Ref.~\cite{Green:1987cr}.

Recalling that
\beq
H^{p,q}(X) \simeq H^q(X, \wedge^p T^\star X) \ ,
\eeq
where $T^\star X$ is the cotangent bundle of $X$, we can write the desired cohomologies as
\beq\label{h11h21}
H^{1,1}(X) = H^1(M, T^\star X), \qquad
H^{2,1}(X) \simeq H^{1,2}(X) = H^2(X, T^\star X) \simeq H^1(X, TX) \ .
\eeq
In the second part of the above expression, we have used Serre duality, \eref{serre},
to establish the isomorphism between $H^2(X, T^\star X)$ and $H^1(X, TX)$.

We can therefore concentrate on the computing $H^1(X, TX)$. We invoke
the Euler sequence which states that, for an embedding of $X$ into an ambient space $\cA$, there is a short exact
sequence
\beq
\sseq{TX}{T\cA|_X}{\cN |_X}\ ,
\eeq
where $\cN$ is the normal bundle of $X$ in $\cA$ and $T\cA$ is the the tangent bundle of $\cA$. The bar and the subscript, $X$, denotes restriction of the bundle to the Calabi-Yau manifold $X$. This induces a long exact sequence in cohomology as
\beq
\ba{cccccccc}
0 
& \to & H^0(X, TX) & \to & H^0(X, T\cA |_X) & \to & H^0(X, \cN |_X) & \to
\\
& \to & H^1(X, TX) & \stackrel{d}{\to} 
  & H^1(X, T\cA |_X) & \to & H^1(X, \cN |_X) & \to \\
& \to & H^2(X, TX) & \to & \ldots &&&
\ea
\eeq
Since $X$ is a Calabi-Yau manifolds it follows that $H^0(X, TX) = H^{1,3}(X) = 0$. Using this,
the relations \eref{h11h21}, and the fact that $\rk(d) = 0$ (see Eq.~(6.1) of Ref.~\cite{Green:1987cr}), we have the short exact sequence
\beq
\sseq{H^0(X, T\cA |_X)}{H^0(X, \cN |_X)}{H^{2,1}(X)} \ ,
\eeq
and, consequently,
\beq\label{h21}
h^{2,1}(X) = h^0(X, \cN |_X) - h^0(X, T\cA |_X) \ .
\eeq

\subsubsection{Hodge Number Obstructions}
Making use of the essential techniques of Leray tableaux and Koszul resolutions, one can, in principle compute the two terms in Eq.~\eqref{h21} and, hence, obtains the Hodge numbers of complete intersection 3-folds. However, direct calculation shows that one quickly encounters certain obstructions to the computation which will naturally divide our set of 7890 configurations.
\paragraph{Trivial Direct Products}
First of all, we recognize that there are trivial cases in the list,
comprising of CICYs which are simply direct products of
lower-dimensional Calabi-Yau manifolds, viz., $K3 \times T^2$ and
$T^6$. These generically have reduced holonomy and we shall not
consider them. The identifiers for these are 31-52, a total of 22 cases. Therefore, our list is
immediately reduced to be of length 7868.
\paragraph{Normal Bundle and Obstructions}
The Leray $E_1^{j,k}(\cN_X)$ tableaux is readily established for the normal bundle
$\cN_X$ according to \eref{bott} and \eref{kunneth1}. It
turns out that if there exists $j \le j'$ in $[-K,0]$ such
that
\beq
E^{j,j}_1(\cN |_X) \ne  0 \mbox{ and } E^{j',-j'}_1 (\cN |_X)\ne 0 \ ,
\eeq
then, the spectral sequence cannot be iterated to obtain $E_\infty$ without the knowledge of the ranks of some maps.
Such a case, which we call ``normal bundle obstructed'' needs to be addressed separately \cite{Green:1987cr}.
For all remaining cases, the Leray spectral sequence actually terminates at $E_1$ and we can
read off the required cohomology as \cite{Green:1987cr}:
\beq\label{h0N_M}
h^0(X, \cN |_X) = \sum_{j=0}^K e_1^{j,j}(\cN |_X) + 
  \sum_{j=1}^K \sum_{l=0}^{j-1} (-1)^{j+l} e^{l,j}_1(\cN |_X) \ .
\eeq
In the above, we have used, and shall henceforth adopt, the notation
that $h^j$ is the dimension of the cohomology group $H^j$,
$e^{j,k}_r$ is the dimension of $E^{j,k}_r$.

Now, we find a total of 12 normal bundle obstructed cases, namely the CICYs
with identifiers 1443, 1877, 2569, 2980, 3747, 4228, 4448, 4757, 6174, 6229, 7236 and 7243. For these, Ref.~\cite{Green:1987cr} gives a rule to replace the configuration matrix by an isomorphic one which does not have a normal bundle obstruction. To this equivalent configuration, Eq.~\eref{h0N_M} can then be directly applied.

\paragraph{Tangent Bundle and Obstructions}

Like the normal bundle spectral sequence the tangent bundle
spectral sequence can, in general, be obstructed, that is, one cannot compute $E_\infty$
without knowledge of specific maps. However, for the case of compete
intersection calabi-yau manifolds we are saved from this difficulty by
several useful results. 

The first such result is that for a particular class of configurations (those
without a decomposing $(n-1)$-leg, see Ref.~\cite{Green:1987cr} for a description of the
dot/leg diagrams and notation), $E_{1}^{q+k,k}(V)$ vanishes for $q\geq n-1$
for any bundle $V$ on $X.$ It turns out that if a diagram representing a
Calabi-Yau $3$-fold has no decomposing $1$-legs, $H^{1}(X,T\mathcal{A})$ vanishes
and no decomposing $2$-legs implies that $H^{2}(X,N)=0$ so that the sequences%
\begin{align}
0  & \rightarrow H^{0}(X,T\mathcal{A})\rightarrow H^{0}(X,N_X)\rightarrow
H^{1}(X,TX)\rightarrow 0\nonumber\\
0  & \rightarrow H^{1}(X,N_X)\rightarrow H^{2}(X,TX)\rightarrow H^{2}%
(X,T\mathcal{A})\rightarrow0
\end{align}
are exact \cite{Green:1987cr}.

For $3$-folds with decomposing $1$-legs the hodge numbers can be computed by
relying on the classification of complex surfaces (see Eq.~(2.4)in Ref.~\cite{Green:1987cr}). Simple formulas for these
Hodge numbers in terms of sub-diagrams were found in \cite{Green:1987cr}. 
For the bulk of cases, however, the diagrams have no decomposing $1$-legs.

Further, it can be shown that an $n$-fold configuration with the property of a
decomposing $(n-1)$-leg is equivalent to another one with no decomposing
$(n-1)$-leg \cite{Green:1987cr}. So in analysing configurations representing Calabi-Yau $3$-folds it is sufficient to look only at configurations with no decomposing 2-legs. This leads to the following structure
\begin{align}
E_{1}^{0,0}(T\mathcal{A})  & =\bigoplus_{r=1}^{m}H^{0}(\mathbb{P}_{r}^{n_{r}},T(\mathbb{P}_{r}^{n_{r}}));
\nn \\
E_{1}^{K+2,K}(T\mathcal{A})  & \approx C^{m}; 
\nn \\
E_{1}^{q+k,k}(T\mathcal{A)}  & \backepsilon H^{0}(\mathbb{P}_{r}^{n_{r}},
	T(\mathbb{P}_{r}^{n_{r}})\otimes h_{r}^{-1})\approx C^{n_{r}+1},
\quad
\forall\{A,r : \sum_{a\in S}q_{a}^{r}=1,k=\left|  S \right|  +1\}\nonumber
\end{align}
where $\left| S \right|$ denotes the cardinality of $S$, the set of indices
labeling a subset of constraints which act only in a $(q+k)$-dimensional
factor of the ambient space $\cA$. With these results in hand, we can compute the
Hodge numbers of $X$.

\subsection{Redundancy in the CICY list}
It is worth observing that the $7890$ CICYs which appear in the original list are presumably not all topologically distinct~ \cite{philip}. This is a relatively new observation and should be pointed out. 

Wall's theorem (see, for example, Ref.~\cite{hubsch}) states that for real
six-folds, the intersection form and the second Potryagin class suffice
to distinguish non-isormophism. Though for complex threefolds, these
are not enough, the two quantities are good indicators (and will be enough to distinguish our heterotic models). Therefore, we propose a simple check for redundancy. We compare the basic topological invariants
Hodge numbers $h^{1,1}(X)$, $h^{2,1}(X)$, second Chern class $c_{2r}(TX)$, and intersection numbers $d_{rst}$, and identify any two CICYs with identical sets, up to permutation in the indices $r,s$.

Upon implementing such a
scan one finds, of the 7890 in the original list, that 
there are 378 sets of redundancies, consisting of equivalent pairs,
triples, or
even n-tuples for n as large as 6. These are expected to have
isomorphism. In all, 813 manifolds are involved; taking one
representative from each of the 378 sets, a total of 435 CICY seem
redundant. Throughout the rest of the paper, however, we will adhere
to the original identifier names of the manifolds to avoid confusion
and shall point out explicitly, where necessary, the equivalences.



\end{document}